\documentclass[epj]{svjour}
\usepackage{graphicx}
\usepackage{color}

\def\mb#1{\setbox0=\hbox{$#1$}\kern-.025em\copy0\kern-\wd0
\kern-0.05em\copy0\kern-\wd0\kern-.025em\raise.0233em\box0}

\begin{document}
   \title{Dynamical and thermodynamical stability of isothermal
   distributions in the HMF model}

 \author{P.H. Chavanis}

\institute{$^1$ Laboratoire de Physique Th\'eorique (IRSAMC), CNRS and UPS, Universit\'e de Toulouse, F-31062 Toulouse, France\\
\email{chavanis@irsamc.ups-tlse.fr}
}

\titlerunning{Dynamical and thermodynamical stability of isothermal
   distributions in the HMF model}

   \date{To be included later }

   \abstract{We provide a new derivation of the conditions of
   dynamical and thermodynamical stability of homogeneous and
   inhomogeneous isothermal distributions in the
   Hamiltonian Mean Field (HMF) model. This proof completes the original thermodynamical 
   approach of Inagaki [Prog. Theor. Phys. {\bf 90}, 557 (1993)]. Our formalism, based on
   variational principles, is simple and the method can be applied to
   more general situations. For example, it can be used to settle the
   dynamical stability of polytropic distributions with respect to the
   Vlasov equation [Chavanis \& Campa, arXiv:1001.2109]. For
   isothermal distributions, the calculations can be performed fully
   analytically, providing therefore a clear illustration of the
   method.   }

   \maketitle
%

\section{Introduction}
\label{sec_model}

The statistical mechanics of systems with long-range interactions has
recently been the object of an intense activity
\cite{houches,assisebook,oxford,cdr}. A system with long-range
interactions is characterized by a binary potential $u(r)$ which
decreases at large distances $r$ slower than $r^{-d}$ where $d$ is the
dimension of space. Systems with long-range interactions are numerous
in nature and include for example self-gravitating systems,
two-dimensional vortices, bacterial populations experiencing
chemotaxis, neutral and non neutral plasmas, wave-particle systems,
free-electron lasers,...  These systems exhibit a very interesting
dynamics and thermodynamics. On a dynamical point of view, they
display robust and long-lived quasi-stationary states (QSSs) that are
non-Boltzmannian. These distributions are stable steady states of the
Vlasov equation on the coarse-grained scale resulting from a process
of violent relaxation \cite{lb}.  On a thermodynamical point of view,
their Boltzmannian statistical equilibrium states can display numerous
types of phase transitions due to ensembles inequivalence
\cite{ellis,bb}. Ensembles inequivalence is generic for systems with
long-range interactions, as first evidenced in astrophysics (see reviews in
\cite{paddy,katzrev,ijmpb}), but not compulsory.

A simple toy model of systems with long-range interactions, called the
Hamiltonian Mean Field (HMF) model, has received a particular
attention.  This model consists of $N$ particles of unit mass moving
on a ring and interacting via a cosine potential $u=N^{-1}
\cos(\theta_i-\theta_j)$ where $\theta_i$ denotes the angle that makes
particle $i$ with an axis of reference. This model was first
introduced by Messer \& Spohn \cite{ms} and called the cosine model. It was
reintroduced independently ten years later by different groups of
researchers \cite{kk,ik,inagaki,pichon,ar} and was very much studied since then (see a short
 history in \cite{cvb}). In particular, the
seminal paper of Antoni \& Ruffo \cite{ar} inspired many works on the subject.

The HMF model exhibits two successive types of relaxation. For short
timescales, the distribution function is governed by the Vlasov
equation that ignores correlations (or ``collisions'') between
particles. In this regime, the system can experience a collisionless
violent relaxation towards a steady state of the Vlasov equation on
the coarse-grained scale. In principle, this quasistationary state
(QSS) can be predicted by using Lynden-Bell's statistical theory of
violent relaxation
\cite{epjb,precommun,anto,proc,staniscia1}. However, this prediction may fail because
collisionless relaxation can be incomplete
\cite{staniscia1,incomplete,cc}. Since the Vlasov equation admits an
infinite number of steady solutions, the prediction of the QSS
actually reached by the system in case of incomplete relaxation is
very difficult, or even impossible. Nevertheless, it can be useful to
derive general stability criteria in order to determine which
distributions are stable or unstable with respect to the Vlasov
equation.  This is the problem of {\it
Vlasov dynamical stability}
\cite{ik,pichon,ar,cvb,cc,yamaguchi,choi,jain,cd,campac}.  Of course, only stable states are relevant to characterize
QSSs in the context of violent relaxation.  On a longer timescale, the
system is expected to achieve a Boltzmannian statistical equilibrium
state due to the development of correlations between particles (finite
$N$ effects or graininess). This statistical equilibrium state
corresponds to the distribution that maximizes the Boltzmann entropy
at fixed mass and energy. Only global or local entropy maxima are
relevant (minima or saddle points must be discarded). This is the
problem of {\it thermodynamical stability}
\cite{ms,inagaki,pichon,ar,cvb,largedev}. These problems of dynamical
and thermodynamical stability have been investigated in the past using
different methods that we shall briefly review in this introduction.

Let us first discuss the problem of thermodynamical stability.  Messer
\& Spohn \cite{ms} considered a potential energy of the form
$U=N^{-1}\sum_{i<j}V(x_i,x_j)$ and proved rigorously that the mean
field approximation is exact for $N\rightarrow +\infty$ and that the
statistical equilibrium state in the canonical ensemble corresponds to
the global minimum of free energy $F[f]$ at fixed mass
$M$. Considering specifically the cosine model, they showed that it
displays a second order phase transition from a homogeneous phase to a
clustered phase below a critical temperature $T_c$.  Inagaki
\cite{inagaki} studied the thermodynamical stability of the modified
Konishi-Kaneko \cite{kk} system in the microcanonical ensemble by
considering the maximization of entropy $S[f]$ at fixed mass $M$ and
energy $E$. By using the theory of Poincar\'e on linear series of
equilibria (see, e.g., \cite{ijmpb,katz} for details), or by studying
the sign of the second order variations of entropy (for the
homogeneous phase), he showed that the system displays a
microcanonical second order phase transition at a critical energy
$E_c$, corresponding to the critical temperature
$T_c$\footnote{Although not explicitly mentioned by Inagaki
\cite{inagaki}, it is clear that a direct application of the
Poincar\'e theorem shows that the statistical ensembles are equivalent
for this model.}.  Similar results were obtained by Pichon
\cite{pichon} who showed in addition that this type of phase
transitions could explain the formation of bars in disk
galaxies. Antoni \& Ruffo \cite{ar} studied the statistical
equilibrium state of the HMF model in the canonical ensemble directly
from the partition function. They simplified it by using the
Hubbard-Stratonovich transformation and the saddle point approximation
valid for $N\rightarrow +\infty$. They evidenced a second order phase
transition at $T=T_c$ in the canonical ensemble and performed
numerical simulations in the microcanonical ensemble. These
simulations show a discrepancy with the theoretical results close to
the transition energy $E_c$, but this discrepancy is not due to
ensembles inequivalence but to nonequilibrium effects
\cite{ar,cc,lrt}. More recently, Barr\'e {\it et al.} \cite{largedev}
studied the statistical mechanics of the HMF model by applying large
deviation technics. They confirmed the existence of microcanonical and
canonical second order phase transitions and the equivalence of
statistical ensembles. Finally, Chavanis {\it et al.} \cite{cvb}
pursued the thermodynamical approach of Inagaki \cite{inagaki} based
on variational principles. In particular, they reduced the stability
problem to the study of an eigenvalue equation and solved this
eigenvalue equation numerically for any energy and analytically close
to the critical point $(E_c,T_c)$. They proved by this method that the
statistical ensembles are equivalent and that the homogeneous states
are stable for $E>E_c$ (or $T>T_c$) and unstable for $E<E_c$ (or
$T<T_c$). Below that critical energy or critical temperature, they are
replaced by inhomogeneous states that are always stable. This method
has the advantage of showing which type of perturbation is able to
trigger the instability of the homogeneous phase below $E_c$ or $T_c$.

Let us now review the results concerning the dynamical stability of
steady states of the Vlasov equation in the context of the HMF
model. We first consider the linear dynamical stability problem.
Inagaki \& Konishi \cite{ik} and Pichon \cite{pichon} studied the
linear dynamical stability of the isothermal (Maxwell) distribution
with respect to the Vlasov equation.  They considered the homogeneous
phase and derived the dispersion relation using the methods of plasma
physics and stellar dynamics. They showed that the system becomes
dynamically unstable below the critical temperature $T_c$ or the
critical energy $E_c$ (the same as the ones arising in the
thermodynamical approach) leading to an instability similar to the
Jeans instability in self-gravitating systems. They showed that only
the modes $n=\pm 1$ grow, leading to the formation of a single
cluster.  Inagaki \& Konishi \cite{ik} also compared their theoretical
results with direct numerical simulations, finding a good agreement.
Antoni
\& Ruffo \cite{ar} studied the linear dynamical stability of the spatially homogeneous waterbag
distribution and determined the critical temperature $T_c'$ and the
critical energy $E_c'$ marking the separation between stable and
unstable states. More recently, Choi \& Choi \cite{choi}, Chavanis
{\it et al.} \cite{cvb} and Jain {\it et al.} \cite{jain} completed
these studies and obtained explicit expressions for the growth rate
and pulsation period of isothermal, polytropic and waterbag
distributions. On the other hand, Chavanis \& Delfini \cite{cd}
performed an exhaustive study of the linear dynamical stability of the
HMF model by using the Nyquist method. They considered various types
of distributions (single and double humped) and derived general
stability criteria and stability diagrams for both attractive
(ferromagnetic) and repulsive (antiferromagnetic) interactions.

Let us finally review the results concerning the formal nonlinear
dynamical stability of a steady state of the Vlasov equation. A
distribution function is said to be formally stable \cite{holm} if it
is a local minimum or a local maximum of an energy-Casimir functional
(i.e. the second variations of the energy-Casimir functional are
positive definite or negative definite). Formal stability implies
linear stability, but the converse is wrong in general. Yamaguchi {\it
et al.} \cite{yamaguchi} derived a necessary and sufficient condition
of formal nonlinear dynamical stability for spatially homogeneous
distribution functions of the HMF model. They observed that linear
stability and formal stability criteria coincide in that
case. Chavanis {\it et al.} \cite{cvb} reconsidered the formal
stability problem on a new angle (see also \cite{cd,assise} for more
details) which can be extended to more general
situations\footnote{This method is related to the Antonov first law in
astrophysics \cite{bt,aaantonov}.}. They showed that the variational
problem for the distribution function $f(\theta,v)$ is equivalent to a
simpler variational problem for the density $\rho(\theta)$.  This
equivalence is valid for both homogeneous and inhomogeneous
distributions. For spatially homogeneous distributions, they showed
that the condition of formal stability can be written as a condition
on the velocity of sound in the corresponding barotropic gas. This is
equivalent to the criterion derived by Yamaguchi {\it et al.}
\cite{yamaguchi} but expressed in a different manner\footnote{These
criteria have been used in
\cite{cvb,yamaguchi} to determine the formal stability of isothermal
and polytropic distributions and in \cite{epjb} to determine the
formal stability of the Lynden-Bell (or Fermi-Dirac)
distribution.}. For spatially inhomogeneous distributions, they
reduced the formal stability problem to the study of an eigenvalue
equation.  Chavanis \& Delfini \cite{cd} showed that this eigenvalue
equation can be solved analytically at the point of neutral
stability, providing therefore an explicit condition to locate this
point in the series of equilibria. They also generalized the preceding
results to arbitrary potentials of interaction $u({\bf r},{\bf r}')$
and discussed more refined criteria of dynamical stability that are
obtained by conserving a larger class of Vlasov invariants.

Recently, Campa \& Chavanis \cite{campac} derived a very general
criterion of linear dynamical stability valid for homogeneous and
inhomogeneous distributions. They also derived sufficient conditions
of linear and formal dynamical stability that are weaker than the
general criterion but more explicit. For spatially homogeneous
distributions, they proved that the criteria of formal and linear
stability coincide. As a by-product, their results also return the
thermodynamical stability criteria obtained previously by different
methods.

Let us finally mention that the Vlasov dynamical stability of polytropic distributions has been studied by Chavanis \& Campa \cite{cc} by plotting the series of equilibria and using the Poincar\'e criterion. On the other hand, the  Vlasov dynamical stability and the thermodynamical stability of the Lynden-Bell distributions have been studied by Antoniazzi {\it et al.} \cite{anto} and Staniscia {\it et al.} \cite{staniscia1} by solving the variational problem numerically. These studies exhibit a rich phase diagram with several types of phase transitions showing the complexity of the stability problem in the general case.

In this paper, we shall present a new method to determine the
dynamical and thermodynamical stability of isothermal distributions of
the HMF model. The idea is to start from the fundamental variational
problems (\ref{sf1}) and (\ref{ff1}) and transform them into
equivalent but simpler variational problems until a point at which
they can be explicitly solved. This completes the thermodynamical
approach of Inagaki \cite{inagaki} by proving analytically the
stability of the {\it inhomogeneous} phase which was not done in
Inagaki's paper. An interest of our approach is its simplicity and
generality. Indeed, it can be extended to solve Vlasov and
Lynden-Bell stability problems. For example, it has been used recently
to settle the Vlasov dynamical stability of polytropic distributions
(see Sec. 8 of
\cite{cc}). However, in that case, the calculations are less explicit
than for isothermal distributions. It is therefore interesting to
develop the calculations in detail in the case of isothermal
distributions (where they are fully analytical) in order to clearly
illustrate the method. Although we rederive known results in a
different manner, we think that the present approach is interesting
and potentially useful to tackle more general problems. In addition,
our approach not only determines the strict equilibrium state (global
maximum of entropy) but it can also say whether a critical point of
entropy is metastable (local entropy maximum) or unstable (saddle
point of entropy). Although isothermal distributions of the HMF model
do not display metastable states, this information can be valuable in
more general situations \cite{anto,staniscia1,cc} where our method can
be applied.

\section{Series of equilibria}
\label{sec_se}

The HMF model is a system of $N$ particles of unit mass $m=1$ moving on a circle and interacting via a cosine potential. The dynamics of these particles is governed by the Hamilton equations \cite{inagaki,pichon,ar,cvb}:
\begin{eqnarray}
\frac{d\theta_i}{dt}=\frac{\partial H}{\partial v_i}, \qquad \frac{d v_i}{dt}=-\frac{\partial H}{\partial \theta_i},\nonumber\\
H=\frac{1}{2}\sum_{i=1}^{N} v_i^2-\frac{k}{4\pi}\sum_{i\neq j}\cos(\theta_i-\theta_j),
\label{gr1}
\end{eqnarray}
where $\theta_i\in [-\pi,\pi]$ and $-\infty<v_i<+\infty$ denote  the position (angle) and the velocity of particle $i$ and $k$ is the coupling constant (we assume here that $k>0$).
The thermodynamic limit corresponds to $N\rightarrow +\infty$ in such a way that the rescaled energy $\epsilon=8\pi E/kM^2$ remains of order unity. We can take $k\sim 1/N$ which is the Kac prescription. In that case, the energy is extensive, $E/N\sim 1$, but non-additive. For $N\rightarrow +\infty$, the mean field approximation is exact and the $N$-body distribution function  is a product of $N$ one body distributions: $P_{N}(\theta_1,v_1,...,\theta_N,v_N,t)=P_{1}(\theta_1,v_1,t)...P_{1}(\theta_N,v_N,t)$. 

Let us introduce the distribution function $f(\theta,v,t)=NP_{1}(\theta,v,t)$.  For a fixed interval of time and $N\rightarrow +\infty$, the evolution of the distribution function $f(\theta,v,t)$ is governed by the Vlasov equation
\begin{eqnarray}
\label{gr2}
\frac{\partial f}{\partial t}+v\frac{\partial f}{\partial\theta}-\frac{\partial\Phi}{\partial\theta}\frac{\partial f}{\partial v}=0,
\end{eqnarray}
where
\begin{equation}
\Phi(\theta,t)=-\frac{k}{2\pi}\int_{0}^{2\pi}\cos(\theta-\theta')\rho(\theta',t)\, d\theta',\label{gr3}
\end{equation}
is the self-consistent potential generated by the density of particles $\rho(\theta,t)=\int f(\theta,v,t)\, dv$. The mean force acting on a particle located in $\theta$ is $\langle F\rangle=-\partial\Phi/\partial\theta(\theta,t)$. Expanding the cosine function in equation (\ref{gr3}), we obtain
\begin{eqnarray}
\label{se6}
\Phi(\theta,t)=B_x\cos\theta+B_y \sin\theta,
\end{eqnarray}
where
\begin{eqnarray}
\label{se7}
B_x=-\frac{k}{2\pi}\int\rho(\theta,t)\cos\theta\, d\theta,
\end{eqnarray}
\begin{eqnarray}
\label{se8}
B_y=-\frac{k}{2\pi}\int\rho(\theta,t)\sin\theta\, d\theta,
\end{eqnarray}
are proportional to the magnetization (with the opposite sign). The magnetization can be viewed as the order parameter of the HMF model.

Let us introduce the mass
\begin{eqnarray}
\label{gr5}
M[\rho]=\int \rho \, d\theta,
\end{eqnarray}
and the mean field energy
\begin{eqnarray}
\label{gr6}
E[f]=\frac{1}{2}\int f v^2\, d\theta dv+\frac{1}{2}\int \rho\Phi\, d\theta=K+W,
\end{eqnarray}
where $K$ is the kinetic energy and $W$ the potential energy. Using equations (\ref{se6})-(\ref{se8}), the potential energy can be expressed in terms of the magnetization as
\begin{eqnarray}
\label{se16}
W=-\frac{\pi B^2}{k}.
\end{eqnarray}
We also introduce the Boltzmann entropy
\begin{eqnarray}
\label{ff2}
S[f]=-\int f\ln \left (\frac{f}{N}\right )\, d\theta dv,
\end{eqnarray}
and the Boltzmann free energy
\begin{eqnarray}
\label{ff2b}
F[f]=E[f]-T S[f],
\end{eqnarray}
where $T=1/\beta$ is the temperature. In the microcanonical ensemble, the statistical equilibrium state is determined by the maximization problem
\begin{eqnarray}
\label{sf1}
\max_f\left\lbrace S\lbrack f\rbrack\, |\, E\lbrack f\rbrack=E,\, M\lbrack f\rbrack=M\right\rbrace.
\end{eqnarray}
In the canonical ensemble,   the statistical equilibrium state is determined by the minimization problem
\begin{eqnarray}
\label{ff1}
\min_f\left\lbrace F\lbrack f\rbrack\, |\, M\lbrack f\rbrack=M\right\rbrace.
\end{eqnarray}
The Boltzmann entropy functional (\ref{ff2}) and the maximum entropy principle (\ref{sf1}) can be justified from a standard combinatorial analysis (see, e.g., \cite{ijmpb,cvb}). The distribution function $f(\theta,v)$ that is solution of (\ref{sf1}) is the {\it most probable macroscopic state}, i.e. the macrostate that is the most represented at the microscopic level, assuming that the accessible  microstates (with the proper values of mass and energy) are equiprobable (see Appendix \ref{sec_heur}).

We shall first determine the {\it critical points} of these variational problems. This will allow us to set the notations that will be needed in the following. The critical points of the maximization problem (\ref{sf1}) are determined by the variational principle
\begin{eqnarray}
\label{se1}
\delta S-\beta\delta E-\alpha\delta M=0,
\end{eqnarray}
where $\beta=1/T$ and $\alpha$ are Lagrange multipliers associated with the conservation of energy and mass. The critical points of the minimization problem (\ref{ff1}) are determined by the variational principle
\begin{eqnarray}
\label{se2}
\delta F+\alpha T\delta M=0,
\end{eqnarray}
where $\alpha$ is a Lagrange multiplier associated with the conservation of mass. Since $T$ is fixed in the canonical ensemble, it is clear that equation (\ref{se2}) is equivalent to equation (\ref{se1}). Therefore, the optimization problems (\ref{sf1}) and (\ref{ff1}) have the {\it same} critical points. Performing the variations in  equations  (\ref{se1}) and (\ref{se2}), we find that the critical points are given by the mean field Maxwell-Boltzmann distribution
\begin{eqnarray}
\label{se3}
f(\theta,v)=A'\, e^{-\beta\left \lbrack\frac{v^2}{2}+\Phi(\theta)\right \rbrack},
\end{eqnarray}
where $A'=e^{-1-\alpha}$ is a constant. Integrating over the velocity,
we get the mean field Boltzmann distribution
\begin{eqnarray}
\label{se4}
\rho(\theta)=A\, e^{-\beta\Phi(\theta)},
\end{eqnarray}
where $A=(2\pi/\beta)^{1/2}A'$. Using the expression (\ref{se6}) of the potential, the distribution function (\ref{se3}) can be rewritten
\begin{eqnarray}
\label{se9}
f(\theta,v)=A'\, e^{-\beta\left (\frac{v^2}{2}+B_x\cos\theta+B_y\sin\theta\right )}.
\end{eqnarray}
It is convenient to write $B_x=B\cos\phi$ and $B_y=B\sin\phi$ with $B=(B_x^2+B_y^2)^{1/2}$. In that case, the foregoing expression takes the form
\begin{eqnarray}
\label{se10}
f(\theta,v)=A'\, e^{-\beta\left \lbrack\frac{v^2}{2}+B\cos(\theta-\phi)\right \rbrack}.
\end{eqnarray}
The corresponding density profile is
\begin{eqnarray}
\label{se11}
\rho(\theta)=A\, e^{-\beta B\cos(\theta-\phi)}.
\end{eqnarray}
The amplitude $A$ and the magnetization $B$ are determined by substituting equation (\ref{se11}) in equations (\ref{se7}), (\ref{se8}) and (\ref{gr5}). This yields
\begin{equation}
A=\frac{M}{2\pi I_{0}(\beta B)},
\label{se12}
\end{equation}
and
\begin{equation}
\frac{2\pi B}{kM}=\frac{I_{1}(\beta B)}{I_{0}(\beta B)},
\label{se13}
\end{equation}
where $I_n(x)$ is the modified Bessel function of order $n$. Equation (\ref{se13}) determines the magnetization $B$ as a function of the temperature $T$. Then, $A$ is given by equation (\ref{se12}). Note that the critical points are degenerate. There exists an infinity of critical points which differ only by their phase $\phi$, i.e. by the position of the maximum of the density profile. They have the same value of entropy or free energy (see below). In the following, we shall take $\phi=0$ without loss of generality. In that case, $B_x=B$ and $B_y=0$. Then, the distribution function and the density can be written
\begin{eqnarray}
\label{se14}
f(\theta,v)=\left (\frac{\beta}{2\pi}\right )^{1/2} \, \rho({\theta})\, e^{-\beta\frac{v^2}{2}},
\end{eqnarray}
\begin{eqnarray}
\rho(\theta)=\frac{M}{2\pi I_0(\beta B)} e^{-\beta B\cos\theta},
\label{se15}
\end{eqnarray}
where $B$ is determined in terms of $T$ by equation (\ref{se13}). The study of the self-consistency relation (\ref{se13}) is classical: $B=0$ is always solution while solutions with $B\neq 0$ only exist for $T<T_c=kM/4\pi$ \cite{cdr,inagaki,cvb}.

Let us now determine the expressions of the energy, entropy and free energy. For the Maxwell-Boltzmann distribution (\ref{se14}), the kinetic energy is
\begin{eqnarray}
\label{se17}
K=\frac{1}{2}MT.
\end{eqnarray}
Combining this relation with equation (\ref{se16}), we find that the total energy $E=K+W$ is given by
\begin{eqnarray}
\label{se18}
E=\frac{1}{2}MT-\frac{\pi B^2}{k}.
\end{eqnarray}
The series of equilibria giving $T$ as a function of $E$ is determined by equations (\ref{se13}) and (\ref{se18}) by eliminating $B$. The relation that gives the magnetization $B$ as a function of the energy $E$ is determined by equations (\ref{se13}) and (\ref{se18}) by eliminating $T$. Finally, using equations (\ref{ff2}) and (\ref{se14}), the entropy is given by
\begin{eqnarray}
\label{se19}
S=\frac{1}{2}M\ln T-\int\rho\ln\rho\, d\theta,
\end{eqnarray}
up to a term $\frac{1}{2}M+\frac{1}{2}M \ln (2\pi)+M\ln M$. Using equation (\ref{se15}), it can be rewritten
\begin{eqnarray}
\label{se20}
S=\frac{1}{2}M\ln T+M\ln I_0(\beta B)-\frac{2\pi B^2}{kT},
\end{eqnarray}
up to a term $\frac{1}{2}M+\frac{3}{2}M \ln (2\pi)$. The relation between the entropy $S$ and  the energy $E$ is determined by equations (\ref{se20}), (\ref{se18}) and (\ref{se13}). Using equations (\ref{se18}) and (\ref{se20}), the free energy (\ref{ff2b}) is given by
\begin{eqnarray}
\label{se21}
F=\frac{1}{2}M T-\frac{1}{2}MT\ln T-MT\ln I_0(\beta B)+\frac{\pi B^2}{k},
\end{eqnarray}
up to a term $-\frac{1}{2}MT-\frac{3}{2}MT \ln (2\pi)$. The relation between the free energy $F$ and the temperature $T$ is determined by equations (\ref{se21}) and (\ref{se13}).

It is convenient to write these equations in parametric form by introducing the parameter $x=\beta B$. Then, we have
\begin{eqnarray}
\label{se22}
b\equiv \frac{2\pi B}{kM}=\frac{I_1(x)}{I_0(x)},
\end{eqnarray}
\begin{eqnarray}
\label{se23}
\eta\equiv \frac{\beta kM}{4\pi}=\frac{x}{2b(x)},
\end{eqnarray}
\begin{eqnarray}
\label{se24}
\epsilon\equiv \frac{8\pi E}{kM^2}=\frac{1}{\eta(x)}-2b(x)^2,
\end{eqnarray}
\begin{eqnarray}
\label{se25}
s\equiv \frac{S}{M}=-\frac{1}{2}\ln\eta(x)+\ln I_0(x)-2\eta(x)b(x)^2,
\end{eqnarray}
\begin{eqnarray}
\label{se26}
f\equiv \frac{8\pi F}{kM^2}=\epsilon(x)-\frac{2}{\eta(x)}s(x).
\end{eqnarray}
These relations apply to the inhomogeneous states ($b\neq 0$). For the homogeneous states ($b=0$), we have
\begin{eqnarray}
\label{se27}
\epsilon=\frac{1}{\eta},\quad s=\frac{1}{2}\ln\epsilon,\quad f=\frac{1}{\eta}+\frac{1}{\eta}\ln\eta.
\end{eqnarray}
The magnetization $b$ takes values between $0$ and $1$, the inverse temperature $\eta$ between $0$ and $+\infty$ and the energy $\epsilon$ between $\epsilon_{min}=-2$ and $+\infty$. The homogeneous states exist for any $\eta\ge 0$ and for any $\epsilon\ge 0$. The inhomogeneous states exist for any $\eta\ge\eta_c=1$ and any $\epsilon_{min}\le \epsilon\le\epsilon_c=1$. The bifurcation point is located at
\begin{eqnarray}
\label{bif}
\epsilon_c\equiv \frac{8\pi E_c}{kM^2}=1,\qquad \eta_c\equiv \frac{\beta_c kM}{4\pi}=1.
\end{eqnarray}
Close to the bifurcation point ($\eta\rightarrow \eta_c=1^+$, $\epsilon\rightarrow \epsilon_c=1^-$), we get (see Appendix \ref{sec_asy}):
\begin{eqnarray}
\label{se28}
b\simeq\sqrt{2(\eta-1)},\quad b\simeq\sqrt{\frac{2}{5}(1-\epsilon)},\quad \eta\simeq 1+\frac{1}{5}(1-\epsilon),\nonumber\\
\end{eqnarray}
\begin{eqnarray}
\label{se28add}
s \simeq -\frac{1}{2}(1-\epsilon),\quad f\simeq 1-\frac{5}{2}(\eta-1)^2.
\end{eqnarray}
Close to the ground state ($\eta\rightarrow +\infty$, $\epsilon\rightarrow \epsilon_{min}=-2$), we get (see Appendix \ref{sec_asy}):
\begin{eqnarray}
\label{se28b}
b\simeq 1-\frac{1}{4\eta},\quad b\simeq 1-\frac{\epsilon+2}{8},\quad \eta\simeq \frac{2}{\epsilon+2},
\end{eqnarray}
\begin{eqnarray}
\label{se28badd}
s \simeq \ln(\epsilon+2),\quad f\simeq -2+\frac{2}{\eta}\ln\eta. 
\end{eqnarray}

\begin{figure}
\begin{center}
\includegraphics[clip,scale=0.3]{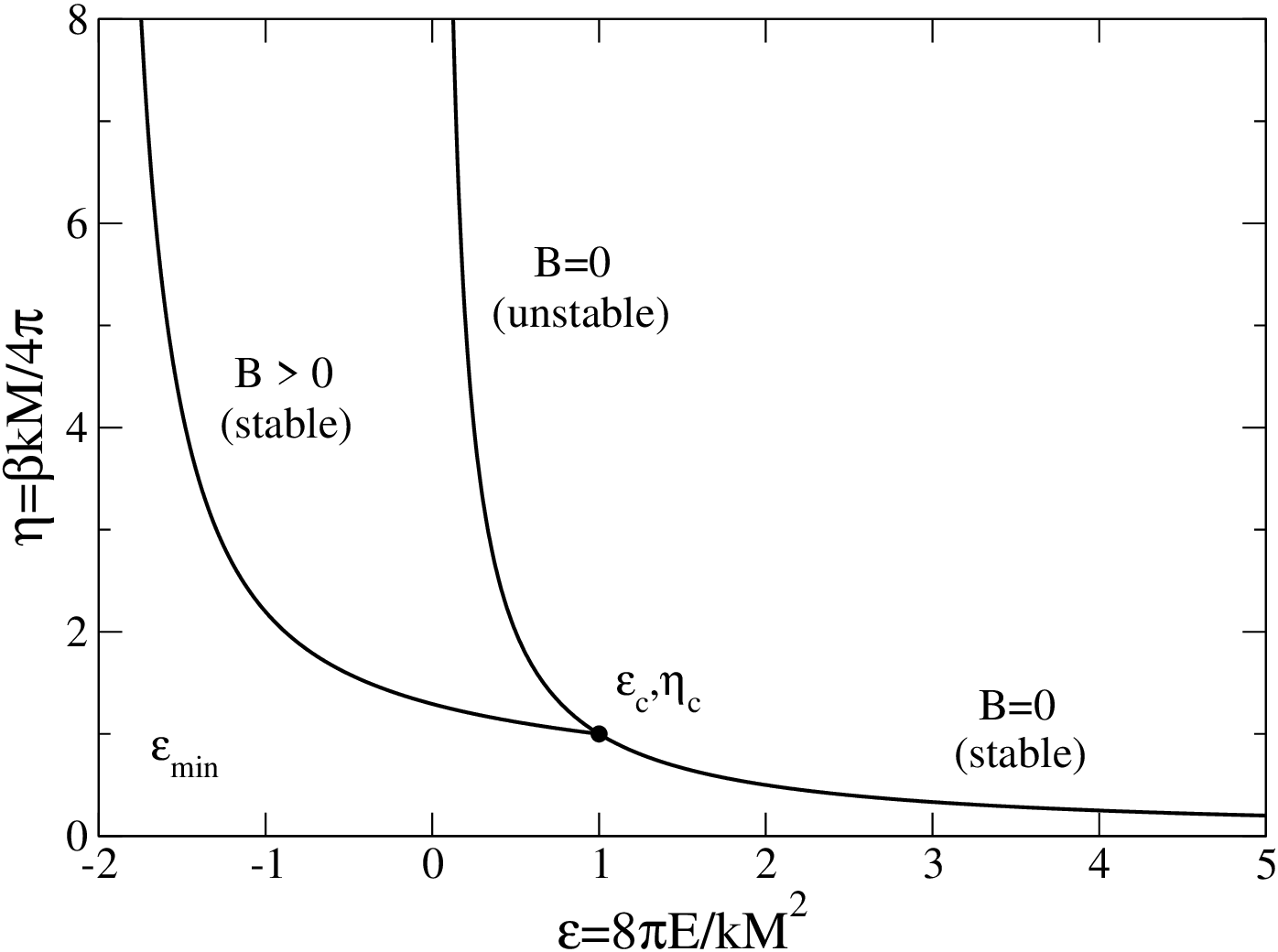}
\caption{Series of equilibria (caloric curve) giving the temperature as a function of energy.}
\label{etaepsilon}
\end{center}
\end{figure}

\begin{figure}
\begin{center}
\includegraphics[clip,scale=0.3]{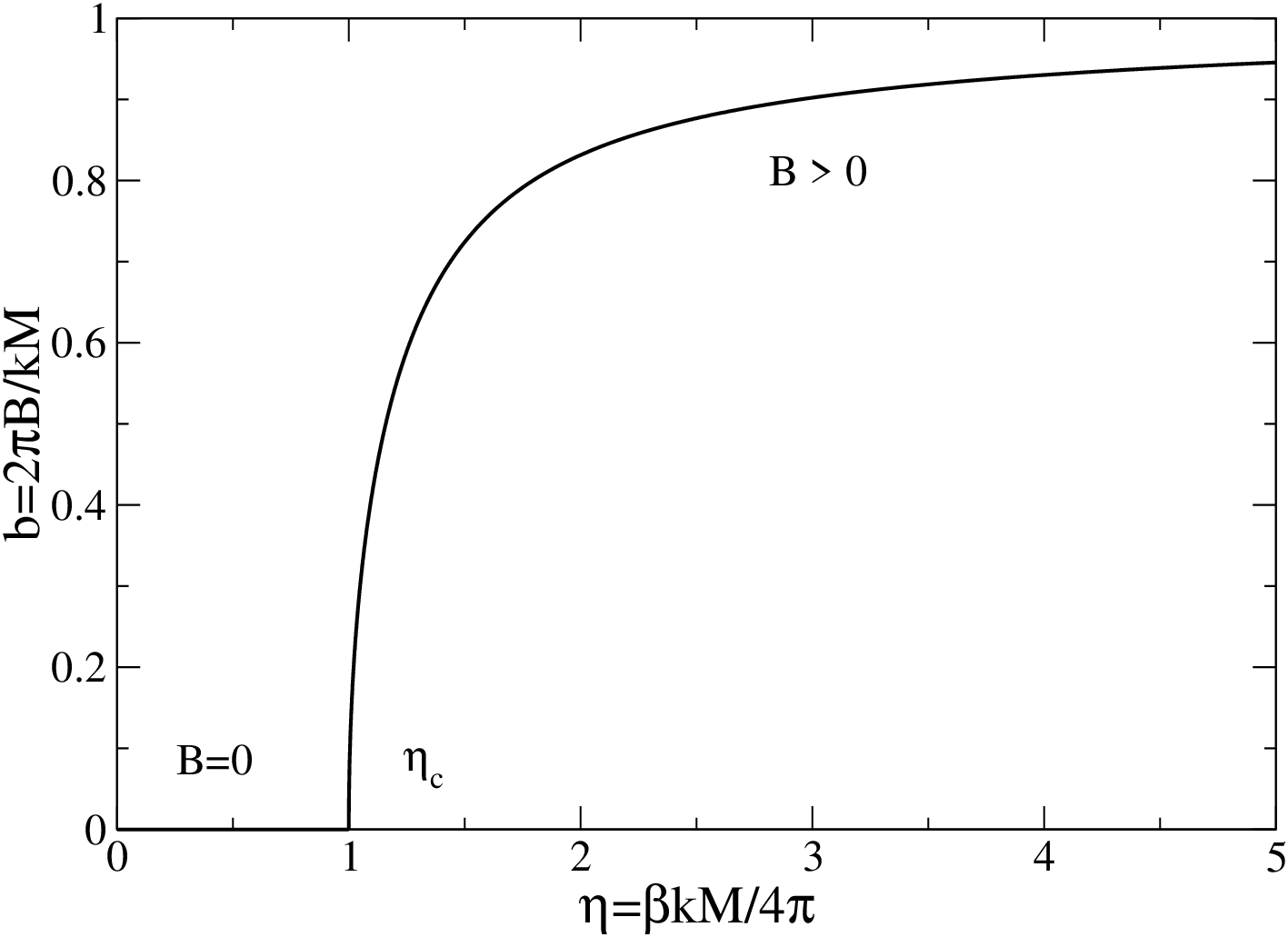}
\caption{Magnetization (order parameter) as a function of temperature.}
\label{beta}
\end{center}
\end{figure}

\begin{figure}
\begin{center}
\includegraphics[clip,scale=0.3]{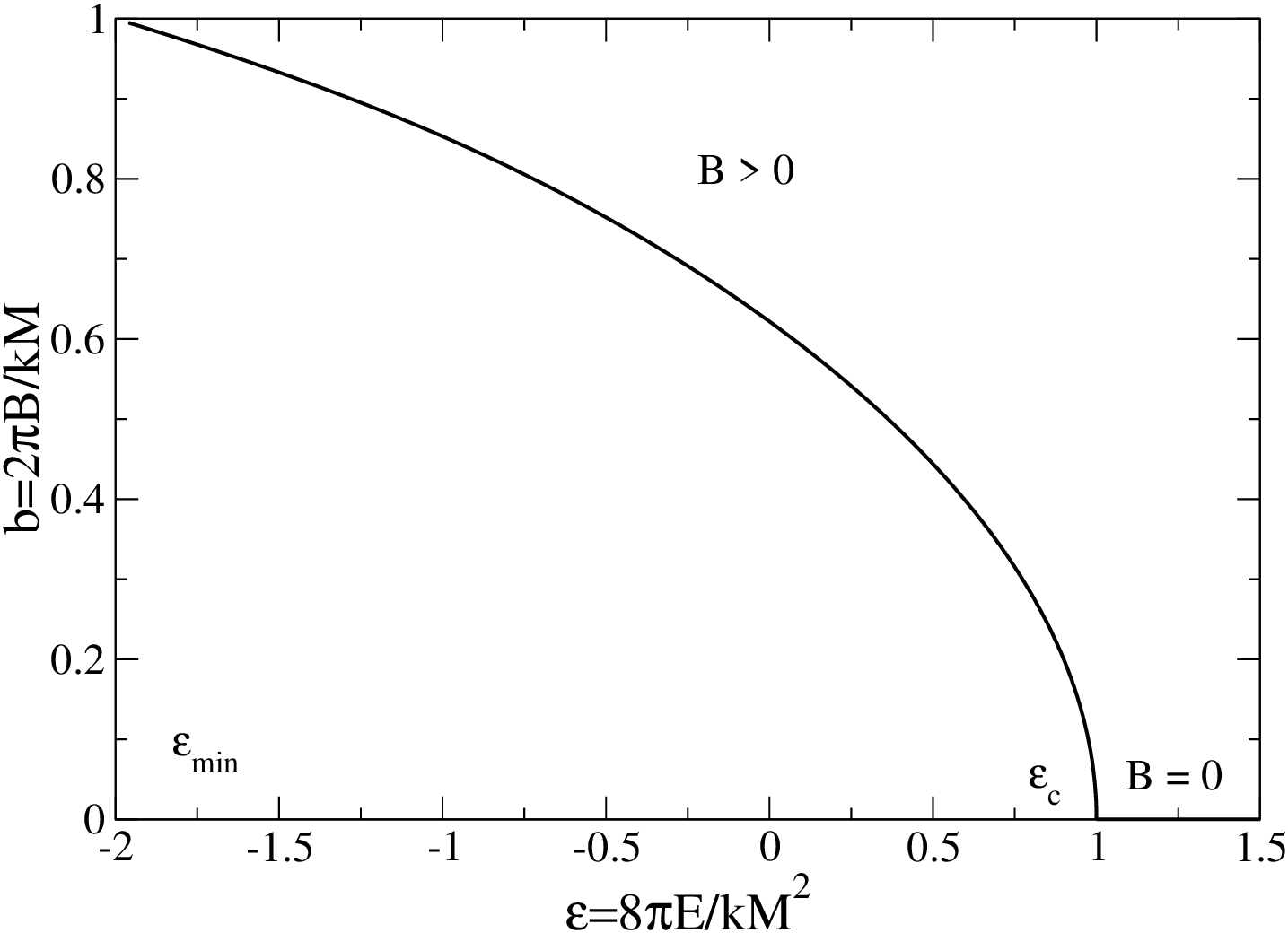}
\caption{Magnetization (order parameter) as a function of energy.}
\label{be}
\end{center}
\end{figure}

\begin{figure}
\begin{center}
\includegraphics[clip,scale=0.3]{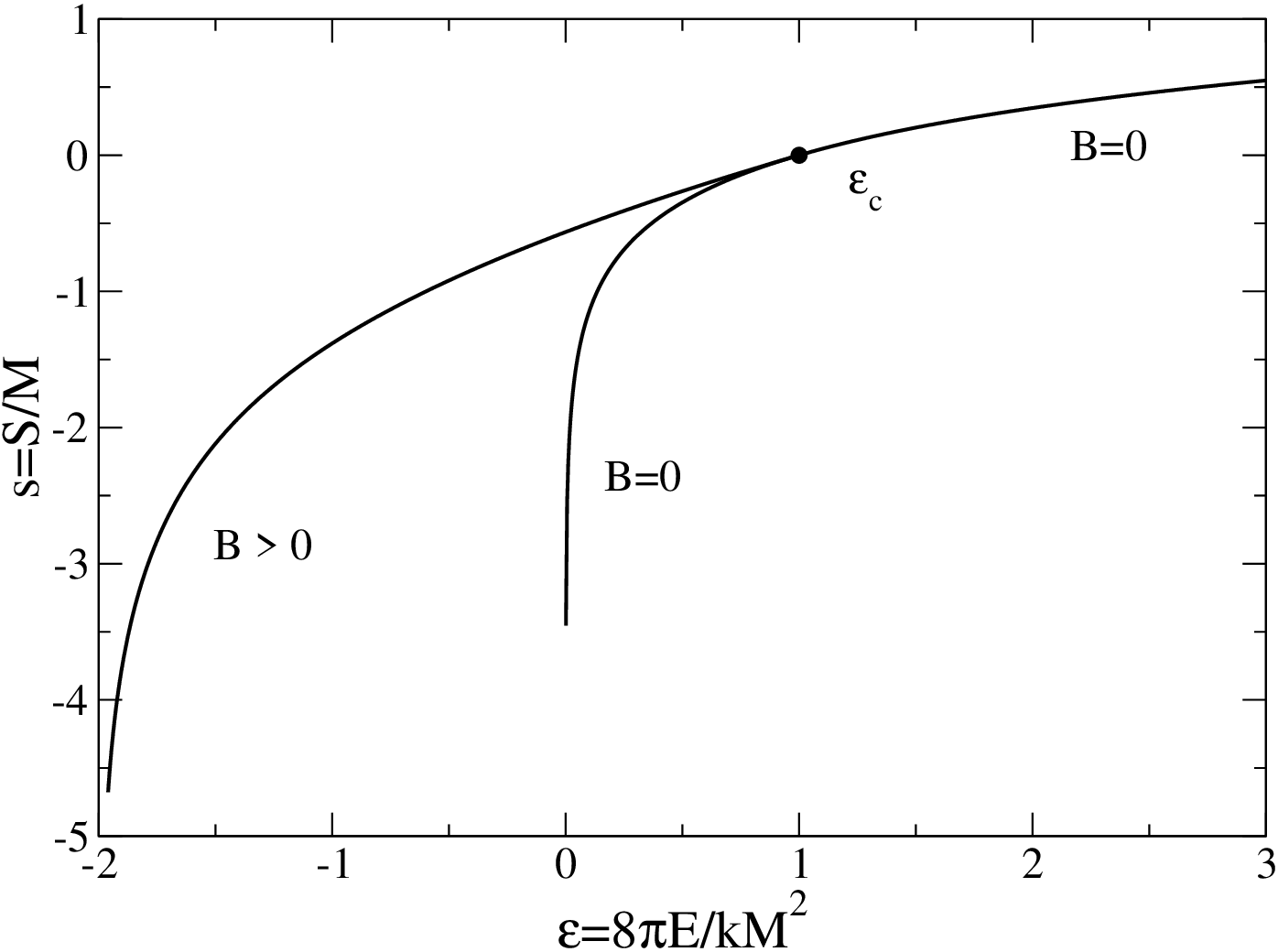}
\caption{Entropy as a function of energy.}
\label{se}
\end{center}
\end{figure}

\begin{figure}
\begin{center}
\includegraphics[clip,scale=0.3]{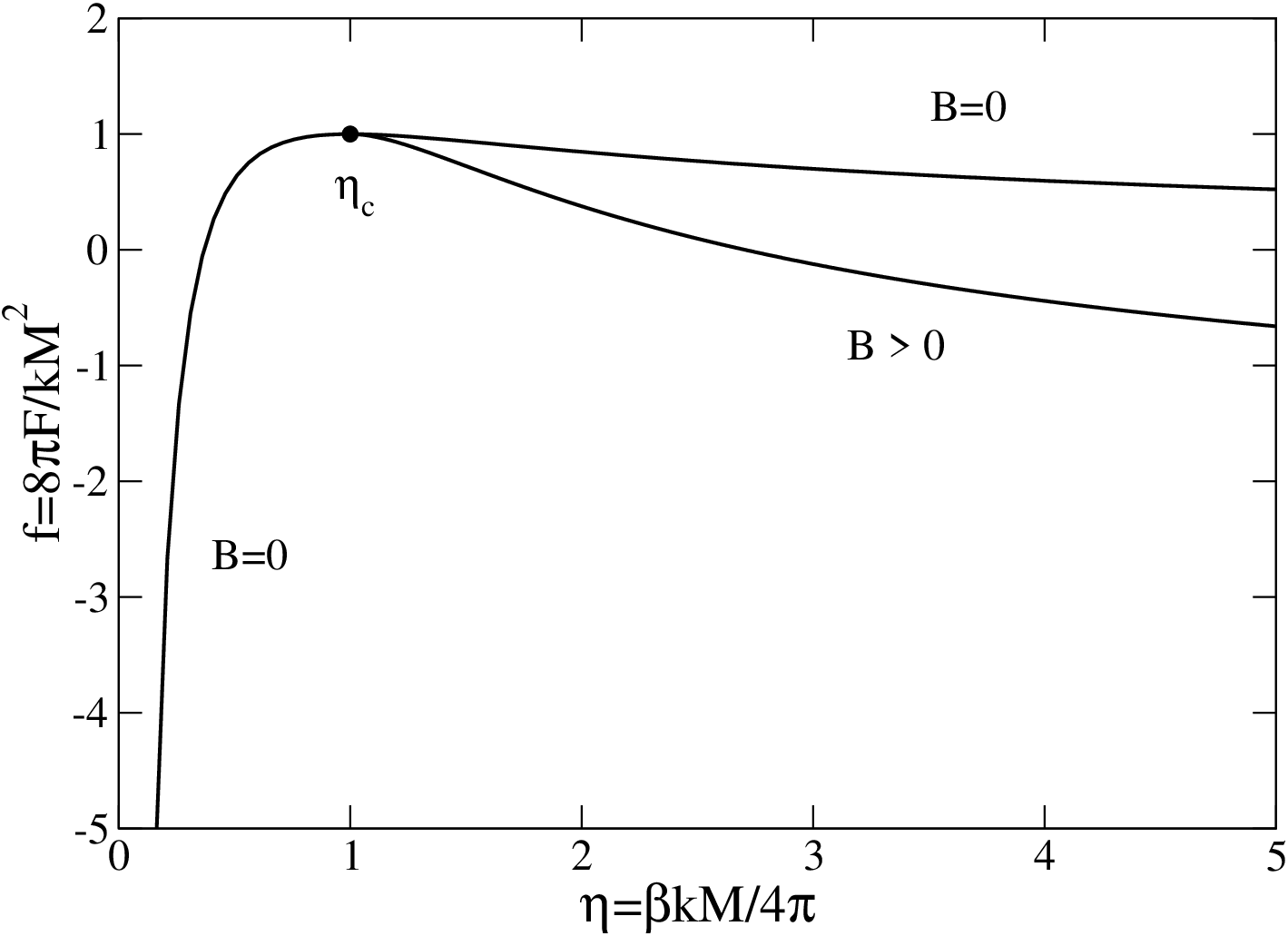}
\caption{Free energy as a function of temperature.}
\label{feta}
\end{center}
\end{figure}

From these relations, we can obtain the curves  $T(E)$, $B(T)$, $B(E)$, $S(E)$ and $F(T)$ in parametric form. These curves are plotted in Figures \ref{etaepsilon}-\ref{feta} for completeness. These results are well-known and they have been derived in many papers \cite{cdr,inagaki,ar,cvb,largedev} using different methods. The present approach, that complements the original approach of Inagaki \cite{inagaki}, is the most direct and the most complete. Indeed, these relations characterize {\it all the critical points} of (\ref{sf1}) and (\ref{ff1}). Now, we must select among these critical points those that are (local) {\it maxima} of $S$ at fixed $E$ and $M$ (microcanonical ensemble) and those that are (local) {\it minima} of $F$ at fixed $M$ (canonical ensemble). This is the object of the next sections.

{\it Remark:} using the Poincar\'e theorem (see, e.g. \cite{ijmpb,katz}), we can directly conclude from the series of equilibria that the homogeneous states are microcanonically (resp. canonically) stable for $E\ge E_c$ (resp. $T\ge T_c$) while they are microcanonically (resp. canonically) unstable for $E\le E_c$ (resp. $T\le T_c$). On the other hand, the inhomogeneous states are always stable. The caloric curve therefore exhibits a microcanonical second order phase transition marked by the discontinuity of $\beta'(E)=S''(E)$ at $E=E_c$  and a canonical second order phase transition marked by the discontinuity of $E'(\beta)=(\beta F)''(\beta)$ at $\beta=\beta_c$. The ensembles are equivalent. In order to illustrate our method, which remains valid in more general situations, we shall however treat both canonical and microcanonical ensembles.

\section{Canonical ensemble}
\label{sec_cano}

\subsection{The functionals $F[f]$ and $F[\rho]$}
\label{sec_ff}

The minimization problem (\ref{ff1}) has several interpretations:

(i) It determines the statistical equilibrium state of the HMF model in the canonical ensemble. In that thermodynamical interpretation $S$ is the Boltzmann entropy, $F$ is the Boltzmann free energy and $T$ is the thermodynamical temperature. The minimization problem (\ref{ff1}) can therefore be interpreted   as a criterion of canonical thermodynamical stability. For isolated systems that evolve at fixed energy, like the HMF model, the canonical ensemble is not physically justified. For such systems, the proper ensemble to consider is the microcanonical ensemble and the statistical equilibrium state is given by (\ref{sf1}). However, the canonical ensemble always provides a {\it sufficient} condition of microcanonical stability \cite{ellis} and it can be useful in that respect\footnote{For systems with short-range interactions, the ensembles are equivalent and we can choose the one that is the most convenient to make the calculations. For systems with long-range interactions, the ensembles may not be equivalent: grand canonical stability implies canonical stability which itself implies microcanonical stability, but the converse is wrong in general.}. Indeed, if a system is canonically stable at the temperature $T$, then it is automatically granted to be microcanonically stable at the corresponding energy $E=E(T)$. Therefore, we can start by this ensemble and consider the microcanonical ensemble only if the canonical ensemble does not cover the whole range of energies. On the other hand, for systems in contact with a thermal bath fixing the temperature, like the Brownian Mean Field (BMF) model \cite{cvb}, the canonical ensemble is the proper ensemble to consider and the statistical equilibrium state is given by (\ref{ff1}).

(ii) It determines a particular steady state of the Vlasov equation that is formally  nonlinearly dynamically stable\footnote{This is a refined condition of formal stability with respect to the usual criterion \cite{holm} since the mass is treated here as a constraint (see \cite{cd,campac} for a more detailed discussion).}. The minimization problem (\ref{ff1}) can therefore be interpreted as a {\it sufficient} condition of dynamical stability. In that dynamical interpretation, $S$ is a particular Casimir (pseudo entropy), $F$ is an energy-Casimir functional (pseudo free energy) and $T$ is a positive constant. It can be convenient to develop a {\it thermodynamical analogy} \cite{assise} to study this dynamical stability problem and use a common vocabulary. In this way, the methods developed in thermodynamics can be applied to the dynamical stability context.

It is shown in Appendix A.2. of \cite{yukawa} that the solution of (\ref{ff1}) is given by
\begin{eqnarray}
\label{fr2}
f(\theta,v)=\left (\frac{\beta}{2\pi}\right )^{1/2} \, \rho({\theta})\, e^{-\beta\frac{v^2}{2}},
\end{eqnarray}
where $\rho(\theta)$ is the solution of
\begin{eqnarray}
\label{fr4}
\min_\rho\left\lbrace F\lbrack \rho\rbrack\, |\, M\lbrack \rho\rbrack=M\right\rbrace,
\end{eqnarray}
where
\begin{eqnarray}
\label{fr3}
F[\rho]=\frac{1}{2}\int\rho\Phi\, d\theta+{T}\int \rho\ln\rho\, d\theta.
\end{eqnarray}
Therefore, the minimization problems (\ref{ff1}) and (\ref{fr4}) are equivalent:
\begin{eqnarray}
\label{eaq}
(\ref{ff1}) \Leftrightarrow (\ref{fr4}).
\end{eqnarray}
This equivalence holds for global and local minimization \cite{yukawa}: (i) $f(\theta,v)$ is the global minimum of (\ref{ff1}) iff $\rho(\theta)$ is the global minimum of (\ref{fr4}) and (ii) $f(\theta,v)$ is a local minimum of (\ref{ff1}) iff $\rho(\theta)$ is a local minimum of (\ref{fr4}). We are therefore led to considering the minimization problem (\ref{fr4}) which is simpler to study since it involves the density $\rho(\theta)$ instead of the distribution function $f(v,\theta)$.

Before that, let us compare the conditions of stability issued from (\ref{ff1}) and (\ref{fr4}). The critical points of (\ref{ff1}) are given by equations (\ref{se14}), (\ref{se15}) and (\ref{se13}) where $\beta$ is prescribed. A critical point of $F[f]$ at fixed mass is a (local) minimum iff
\begin{eqnarray}
\label{ff3}
\delta^2F=\frac{1}{2}\int\delta\rho\delta\Phi\, d\theta+T\int \frac{(\delta f)^2}{2f}\, d\theta dv>0,
\end{eqnarray}
for all perturbations $\delta f$ that do not change the mass: $\delta M=0$. On the other hand, the critical points of (\ref{fr4}) are given by equations (\ref{se15}) and (\ref{se13}) where $\beta$ is prescribed. A critical point of $F[\rho]$ at fixed mass is a (local) minimum iff
\begin{eqnarray}
\label{fr6}
\delta^2F=\frac{1}{2}\int\delta\rho\delta\Phi\, d\theta+T \int \frac{(\delta\rho)^2}{2\rho}\, d\theta> 0,
\end{eqnarray}
for all perturbations $\delta\rho$ that conserve mass: $\delta M=0$. This stability criterion is equivalent to the stability criterion (\ref{ff3}) but it is simpler because it is expressed in terms of the density instead of the distribution function \cite{yukawa}.

{\it Remark 1:} the thermodynamical approach of Messer \& Spohn \cite{ms} in the canonical ensemble directly leads to the minimization problem (\ref{fr4}) for the density, and justifies it rigorously. Our approach recovers this result by another method. It also shows that this minimization problem provides a sufficient condition of dynamical stability with respect to the Vlasov equation.

{\it Remark 2:} considering  the BMF model \cite{cvb}, the minimization problem (\ref{ff1}) determines stable steady states of the mean field Kramers equation and the minimization problem (\ref{fr4}) determines stable steady states of the mean field Smoluchowski equation. According to the equivalence (\ref{eaq}), we conclude that a distribution function $f(\theta,v)$ is a stable steady state of the mean field Kramers equation iff the corresponding density field $\rho(\theta)$ is a stable steady state of the mean field Smoluchowski equation (see \cite{nfp} for a more general statement).

{\it Remark 3:} the equivalence between (\ref{ff1}) and (\ref{fr4}) can be extended to a larger class of functionals of the form $S[f]=-\int C(f)\, d\theta dv$ where $C$ is convex. Such functionals (Casimirs) arise in the Vlasov dynamical stability problem \cite{cd,campac,holm}. We refer to \cite{cvb,assise,nyquistgrav} for a detailed discussion of this equivalence.

\subsection{The function $F(B)$}
\label{sec_cgm}

\subsubsection{Global minimization}
\label{sec_fb}

The equivalence (\ref{eaq}) is valid for an arbitrary potential of interaction $u({\bf r},{\bf r}')$. Now, for the HMF model, the problem can be further simplified. Indeed, the potential energy is given by equation (\ref{se16}) so that  the free energy (\ref{fr3}) can be rewritten
\begin{eqnarray}
\label{fb6}
F[\rho]=-\frac{\pi B^2}{k}+{T}\int \rho\ln\rho\, d\theta.
\end{eqnarray}
We shall first determine the {\it global} minimum of free energy at fixed mass. To that purpose, we shall reduce the minimization problem (\ref{fb6})  to an equivalent but simpler minimization problem.

To solve the minimization problem (\ref{fb6}), we proceed in two steps\footnote{We have used this method in different situations (see, e.g. \cite{assise,aaantonov,yukawa}).}: we first minimize $F[\rho]$ at fixed $M$ {\it and} $B_x$ and $B_y$. Writing the variational principle as
\begin{eqnarray}
\label{fb7}
\delta \left (\int \rho\ln\rho\, d\theta\right )+\alpha\delta M+\mu_x\delta B_x+\mu_y\delta B_y=0,
\end{eqnarray}
we obtain
\begin{eqnarray}
\label{fb8}
\rho_1(\theta)=Ae^{-\lambda_x\cos\theta-\lambda_y\sin\theta},
\end{eqnarray}
where $A=e^{-1-\alpha}$, $\lambda_x=-\frac{k}{2\pi}\mu_x$ and $\lambda_y=-\frac{k}{2\pi}\mu_y$. The Lagrange multipliers are determined by the constraints $M$, $B_x$ and $B_y$. If we write $B_x=B\cos\phi$ and $B_y=B\sin\phi$ then we find that $\lambda_x=\lambda\cos\phi$ and  $\lambda_y=\lambda\sin\phi$ with $\lambda=(\lambda_x^2+\lambda_y^2)^{1/2}$. Equation (\ref{fb8}) can be rewritten
\begin{eqnarray}
\label{fb8b}
\rho_1(\theta)=Ae^{-\lambda\cos(\theta-\phi)}.
\end{eqnarray}
Finally, $A$ and $\lambda$ are determined in terms of $M$ and $B$ through the equations
\begin{equation}
A=\frac{M}{2\pi I_{0}(\lambda)},
\label{fb9}
\end{equation}
and
\begin{equation}
b\equiv \frac{2\pi B}{kM}=\frac{I_{1}(\lambda)}{I_{0}(\lambda)}.
\label{fb10}
\end{equation}
Equation (\ref{fb8b}) is the (unique) global minimum of $F[\rho]$ with the previous constraints since $\delta^2 F=\frac{1}{2}T\int \frac{(\delta \rho)^2}{\rho}\, d\theta> 0$ (the constraints are linear in $\rho$ so that their second variations vanish). Then, we can express the free energy $F[\rho]$ as a function of $B$ by writing $F(B)\equiv F[\rho_1]$. After straightforward calculations, we obtain
\begin{eqnarray}
F(B)=-\frac{\pi B^2}{k}+T \lambda \frac{2\pi B}{k}-MT\ln I_0(\lambda),
\label{fb11}
\end{eqnarray}
where $\lambda(B)$ is given by equation (\ref{fb10}).   Finally, the minimization problem (\ref{fr4})  is equivalent to the minimization problem
\begin{eqnarray}
\label{fb12}
\min_B\left\lbrace F(B)\right\rbrace,
\end{eqnarray}
in the sense that the solution of (\ref{fr4}) is given by equations (\ref{fb8b}), (\ref{fb9}) and (\ref{fb10})  where $B$ is the solution of (\ref{fb12}). Note that the mass constraint is taken into account implicitly in the variational problem (\ref{fb12}). Therefore, (\ref{fr4}) and (\ref{fb12}) are equivalent for global minimization. However, (\ref{fb12}) is much simpler because, for given $T$ and $M$, we just have to determine the minimum of a {\it function} $F(B)$ instead of the minimum of a functional $F[\rho]$  at fixed mass.

Let us therefore study the function $F(B)$ defined by equations (\ref{fb11}) and (\ref{fb10}). Its first derivative is
\begin{eqnarray}
F'(B)=-\frac{2\pi B}{k}+T \lambda\frac{2\pi }{k}+MT\left (\frac{2\pi B}{kM}-\frac{I_0'(\lambda)}{I_0(\lambda)}\right )\frac{d\lambda}{dB}.\nonumber\\
\label{gmc1}
\end{eqnarray}
Using the identity $I'_0(\lambda)=I_1(\lambda)$ and the relation (\ref{fb10}), we see that the term in parenthesis vanishes. Then, we get
\begin{eqnarray}
F'(B)=\frac{2\pi }{k}(T\lambda-B).
\label{gmc2}
\end{eqnarray}
The critical points of $F(B)$, satisfying $F'(B)=0$, correspond therefore to
\begin{eqnarray}
\lambda=x\equiv \beta B.
\label{gmc3}
\end{eqnarray}
Substituting this result in equation (\ref{fb10}), we obtain the self-consistency relation
\begin{equation}
\frac{2\pi B}{kM}=\frac{I_{1}(\beta B)}{I_{0}(\beta B)},
\label{gmc4}
\end{equation}
which determines the magnetization $B$ as a function of the temperature $T$. This returns the results of Sec. \ref{sec_se}.

Now, a critical point of $F(B)$ is a minimum if $F''(B)> 0$ and a maximum if $F''(B)<0$. Differentiating  equation (\ref{gmc2}) with respect to $B$, we find that
\begin{eqnarray}
F''(B)=\frac{2\pi }{k}\left (T\frac{d\lambda}{dB}-1\right ).
\label{gmc5}
\end{eqnarray}
Therefore, a critical point is a minimum if
\begin{eqnarray}
\frac{dB}{d\lambda}< T,
\label{gmc6}
\end{eqnarray}
and a maximum when the inequality is reversed.

\begin{figure}
\begin{center}
\includegraphics[clip,scale=0.3]{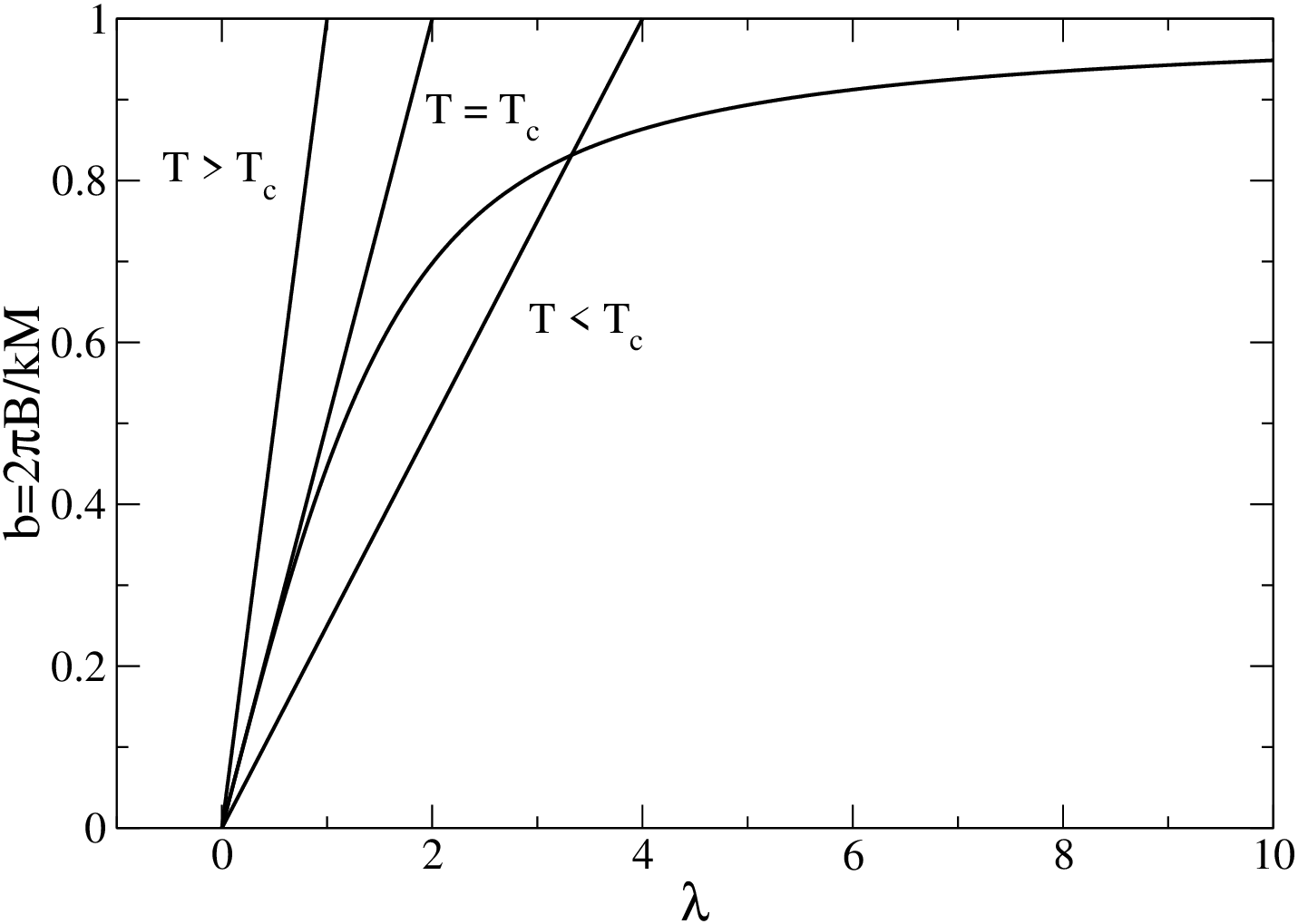}
\caption{Graphical construction determining the critical points of $F(B)$ and their stability. The critical points are determined by the intersection(s) between the curve $b=b(\lambda)$ defined by equation (\ref{fb10}) and the straight line $b=\lambda/(2\eta)$. The critical point is a minimum (resp. maximum) of $F(B)$ if the slope of the curve $b(\lambda)$ at that point is smaller (resp. larger) than the straight line $b=\lambda/(2\eta)$.}
\label{blambda}
\end{center}
\end{figure}

\begin{figure}
\begin{center}
\includegraphics[clip,scale=0.3]{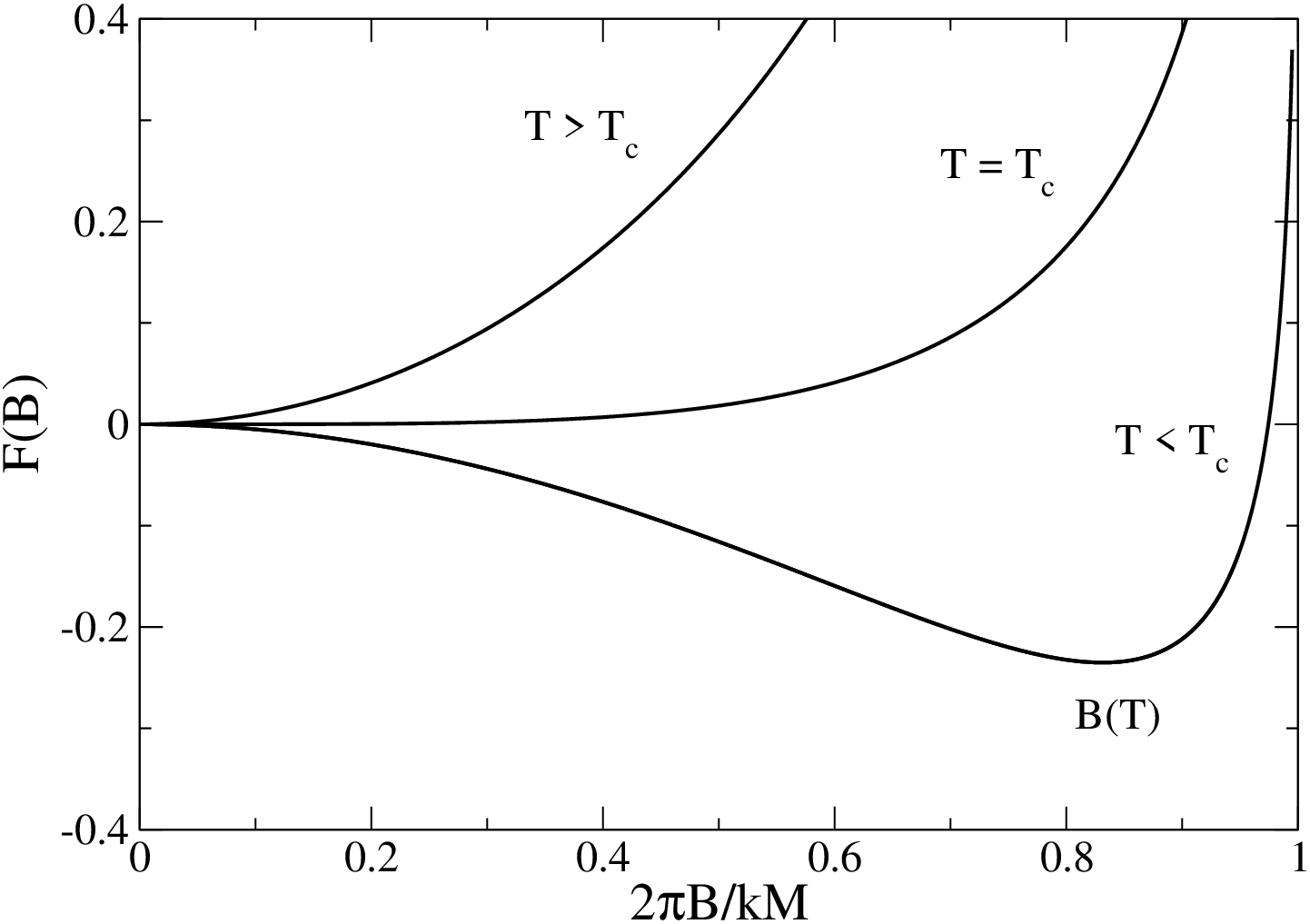}
\caption{Free energy $F(B)$ as a function of the magnetization for a given value of the temperature. For $T>T_c$, this curve has a (unique) global  minimum at $B=0$. For $T<T_c$, this curve has a local maximum at $B=0$ and a global minimum at $B(T)>0$.}
\label{fb}
\end{center}
\end{figure}

We can determine the minimum of the function $F(B)$ by a simple graphical construction. To that purpose,
we plot $b\equiv 2\pi B/kM$ as a function of $\lambda$ according to equation (\ref{fb10}). This is represented in Figure \ref{blambda}. We
find that $2\pi B/kM\rightarrow 1$ for $\lambda\rightarrow
+\infty$.  On the
other hand, $2\pi B/kM\sim \lambda/2$ for $\lambda\rightarrow
0$. Therefore, the magnetization $B$ takes values between
$0$ and $B_{max}=\frac{kM}{2\pi}$.
According to equation (\ref{gmc3}), the critical points of $F(B)$ are determined by the
intersection of this curve with the straight line $2\pi B/kM=(2\pi
T/kM)\lambda$. For $T>T_c\equiv \frac{kM}{4\pi}$, there is a unique
solution $B=0$ corresponding to the homogeneous phase. For $T<T_c$, there are two solutions: a homogeneous
solution $B=0$ and an inhomogeneous solution $B(T)\neq 0$. According to inequality (\ref{gmc6}), a solution is a minimum of $F(B)$  if $d(2\pi
B/kM)/d\lambda<2\pi T/kM$ and a maximum if  $d(2\pi
B/kM)/d\lambda>2\pi T/kM$. Therefore, {\it a critical point of free energy $F(B)$ is a minimum (resp. maximum) if
the slope of the main curve is lower (resp. higher) than the slope of the straight
line at the point of intersection}. From this criterion, we easily conclude that: for $T>T_c$, the homogeneous solution $B=0$ is the global minimum of $F(B)$;  for $T<T_c$, the inhomogeneous solution $B(T)\neq 0$
 is the global minimum of $F(B)$ while the homogeneous solution $B=0$ is a local maximum. Now, using the equivalence between (\ref{fr4}) and (\ref{fb12}) for global minimization, we conclude that the global minimum of the functional $F[\rho]$ at fixed mass is the homogeneous solution for $T>T_c$ and the inhomogeneous solution for $T<T_c$. Furthermore, we will show in the next section that the homogeneous solutions for $T<T_c$ are saddle points of $F[\rho]$ at fixed mass.

To complete our analysis, it can be useful to plot the function $F(B)$ for a prescribed temperature. Using equations (\ref{fb11}) and (\ref{fb10}), the normalized free energy $f\equiv 4\pi F/(kM^2)$ can be expressed in terms of $\lambda$ according to
\begin{eqnarray}
f(\lambda)=-\left (\frac{I_{1}(\lambda)}{I_{0}(\lambda)}\right )^2+\frac{1}{\eta}\left (\lambda\frac{I_1(\lambda)}{I_0(\lambda)}-\ln I_0(\lambda)\right ).
\label{gmc7}
\end{eqnarray}
Eliminating $\lambda$ between the expressions (\ref{gmc7}) and (\ref{fb10}), we obtain the free energy $f(b)$ as a function of the magnetization $b$ for a fixed value of the temperature $T$ (more precisely, for given $\eta$, these equations determine $f(b)$  in a parametric form). For $T>T_c$ and $T<T_c$, this function displays the two behaviors described above, as illustrated in Figure \ref{fb}. For $T\rightarrow T_c$ so that $\lambda,b\rightarrow 0$, we find that the free energy takes the approximate form
\begin{eqnarray}
f(b)\simeq \left (\frac{1}{\eta}-1\right )b^2+\frac{1}{4\eta}b^4, \qquad (\eta\rightarrow 1).
\label{fbza}
\end{eqnarray}
For $\eta>\eta_c=1$, we explicitly check that the minimum satisfying $f'(b)=0$ and $f''(b)>0$ is given by the first relation in equation (\ref{se28}).

{\it Remark 1:} In Appendix \ref{sec_variance}, we plot the second variations of free energy $F''(B(T))$ (related to the variance of the magnetization) as a function of the temperature and recover the previous conditions of stability.

{\it Remark 2:} since the phase $\phi$ does not appear in the function (\ref{fb11}), this means that the inhomogeneous  minima of $F(B)$ for $T<T_c$ are degenerate: there exists  an infinity of minima that only differ in their phases $\phi$.

{\it Remark 3:} since canonical stability implies microcanonical stability \cite{ellis}, and since the series of equilibria $\beta(E)$ is monotonic (see Figure \ref{etaepsilon}), we conclude that the maximum of entropy
at fixed mass and energy is the homogeneous solution for $E>E_c$ and the inhomogeneous solution for $E<E_c$. Since we cover all the accessible range of energies, we conclude that the ensembles are equivalent. We shall, however, treat the microcanonical ensemble specifically in Sec. \ref{sec_micro} since the method can be useful in other contexts where the ensembles are not equivalent (see, e.g. \cite{cc}).

\subsubsection{Local minimization}
\label{sec_lmc}

We shall now show that the minimization problems (\ref{fr4}) and (\ref{fb12}) are also equivalent for {\it local} minimization. To that purpose, we shall relate the second order variations of $F[\rho]$ to the second variations of $F(B)$ by using a suitable decomposition\footnote{This method was previously used  in the context of the statistical mechanics of the 2D Euler \cite{bouchet,euler} and Vlasov \cite{assise} equations.}.

A critical point of (\ref{fr4}) is determined by the variational principle
\begin{eqnarray}
\delta F+\alpha T\delta M=0,
\label{lmc1}
\end{eqnarray}
where $\alpha$ is a Lagrange multiplier accounting for the conservation of mass. This leads to the distribution
\begin{eqnarray}
\rho=\frac{M}{2\pi I_0(\beta B)} e^{-\beta B\cos\theta},
\label{lmc2}
\end{eqnarray}
where we have taken $\phi=0$ without loss of generality (see Sec. \ref{sec_se}). The magnetization $B$ is obtained by substituting equation (\ref{lmc2}) in equation (\ref{se7}) leading to the self-consistency relation (\ref{se13}). Using equations (\ref{fr6}) and (\ref{gr3}), this critical point is a (local) minimum of $F$ at fixed mass iff
\begin{eqnarray}
\delta^2F=-\frac{\pi}{k}((\delta B_x)^2+(\delta B_y)^2)+\frac{1}{2}T\int \frac{(\delta\rho)^2}{\rho}\, d\theta> 0,\nonumber\\
\label{lmc3}
\end{eqnarray}
for all perturbations $\delta\rho$ that conserve mass: $\int \delta\rho\, d\theta=0$. The corresponding variations of magnetization are
\begin{eqnarray}
\delta B_x=-\frac{k}{2\pi}\int \delta\rho\cos\theta\, d\theta,
\label{lmc4a}
\end{eqnarray}
\begin{eqnarray}
\delta B_y=-\frac{k}{2\pi}\int \delta\rho\sin\theta\, d\theta.
\label{lmc4b}
\end{eqnarray}
We can always write the perturbations in the form
\begin{eqnarray}
\delta\rho=\delta\rho_{\|}+\delta\rho_{\perp},
\label{lmc5}
\end{eqnarray}
where $\delta\rho_{\|}=(\mu+\nu_x\cos\theta+\nu_y\sin\theta)\rho$ and  $\delta\rho_{\perp}\equiv \delta\rho-\delta\rho_{\|}$. The second condition ensures that all the perturbations are considered.  We shall now {\it choose}
the constants $\mu$, $\nu_x$ and $\nu_y$ such that
\begin{eqnarray}
\int \delta\rho_{\|}\, d\theta=0,
\label{lmc6}
\end{eqnarray}
\begin{eqnarray}
\delta B_x=-\frac{k}{2\pi}\int
\delta\rho_{\|}\cos\theta\, d\theta,
\label{lmc6b}
\end{eqnarray}
\begin{eqnarray}
\delta B_y=-\frac{k}{2\pi}\int
\delta\rho_{\|}\sin\theta\, d\theta.
\label{lmc6c}
\end{eqnarray}
This implies that
\begin{eqnarray}
\int \delta\rho_{\perp}\,
d\theta=0,
\label{lmc7}
\end{eqnarray}
\begin{eqnarray}
\int
\delta\rho_{\perp}\cos\theta\, d\theta=0, \qquad \int
\delta\rho_{\perp}\sin\theta\, d\theta=0.
\label{lmc7b}
\end{eqnarray}
The conditions (\ref{lmc6}), (\ref{lmc6b}) and (\ref{lmc6c}) lead to the relations
\begin{eqnarray}
\mu M-\frac{2\pi}{k}\nu_x B=0,
\label{lmc8c}
\end{eqnarray}
\begin{eqnarray}
\delta B_x=\mu B-\frac{k\nu_x}{2\pi}I,
\label{lmc9c}
\end{eqnarray}
\begin{eqnarray}
\delta B_y=-\frac{k\nu_y}{2\pi}(M-I),
\label{lmc9bc}
\end{eqnarray}
where we have  defined
\begin{eqnarray}
I\equiv \int \rho\cos^2\theta\, d\theta.
\label{i}
\end{eqnarray}
This forms a system of three algebraic equations that determines the three constants $\mu$, $\nu_x$ and $\nu_y$.
Using the equilibrium distribution (\ref{lmc2}), we find  after simple algebra that
\begin{eqnarray}
I=M\left \lbrack 1-\frac{1}{\beta B}\frac{I_1(\beta B)}{I_0(\beta B)}\right \rbrack,
\label{lmc10}
\end{eqnarray}
where we have used the identity $I_0(x)-I_2(x)=\frac{2}{x}I_1(x)$. Let us define the function ${B}(\lambda)$ by the relation
\begin{equation}
\frac{2\pi {B}(\lambda)}{kM}=\frac{I_{1}(\lambda)}{I_{0}(\lambda)}.
\label{lmc11}
\end{equation}
For $\lambda=\beta B$, according to the self-consistency relation (\ref{se13}), we have ${B}(\beta B)=B$. Differentiating equation (\ref{lmc11}) with respect to $\lambda$, we get
\begin{eqnarray}
\frac{2\pi}{kM}{B}'(\lambda)=\frac{I'_1(\lambda)I_0(\lambda)-I_0'(\lambda)I_1(\lambda)}{I_0(\lambda)^2}.
\label{lmc12}
\end{eqnarray}
Using the identities $I_0'(\lambda)=I_1(\lambda)$, $I_1'(\lambda)=\frac{1}{2}(I_0(\lambda)+I_2(\lambda))$ and $I_0(x)-I_2(x)=\frac{2}{x}I_1(x)$,  and recalling equation (\ref{lmc11}), the foregoing relation can be rewritten
\begin{eqnarray}
\frac{2\pi}{kM}{B}'(\lambda)=1-\frac{1}{\lambda}\frac{I_{1}(\lambda)}{I_{0}(\lambda)}-\frac{4\pi^2 {B}(\lambda)^2}{k^2M^2}.
\label{lmc13}
\end{eqnarray}
Taking $\lambda=\beta B$  and introducing the function $\lambda(B)$, which is the inverse of $B(\lambda)$, we obtain the identity
\begin{eqnarray}
\frac{2\pi}{kM}\frac{1}{\lambda'(B)}=1-\frac{1}{\beta B}\frac{I_{1}(\beta B)}{I_{0}(\beta B)}-\frac{4\pi^2 {B}^2}{k^2M^2}.
\label{lmc13b}
\end{eqnarray}
Comparing  this relation with equation (\ref{lmc10}), we obtain
\begin{eqnarray}
I=\frac{2\pi}{k \lambda'(B)}+\frac{4\pi^2B^2}{k^2M}.
\label{lmc15}
\end{eqnarray}
Solving equations (\ref{lmc8c}), (\ref{lmc9c}) and (\ref{lmc9bc}) for $\mu$, $\nu_x$ and $\nu_y$, and using the result (\ref{lmc15}), we find that
\begin{eqnarray}
\nu_x=-\lambda'(B){\delta B_x},\qquad \mu=-\frac{2\pi B\lambda'(B)}{kM}\delta B_x,
\label{lmc16a}
\end{eqnarray}
\begin{eqnarray}
\nu_y=-\frac{2\pi}{k}\frac{\delta B_y}{M-I}.
\label{lmc16b}
\end{eqnarray}
Therefore, the perturbation $\delta\rho_{\|}$ takes the form
\begin{eqnarray}
\delta\rho_{\|}=- \lambda'(B)\left (\frac{2\pi B}{kM}+\cos\theta\right )\rho(\theta)\delta B_x\nonumber\\
-\frac{2\pi}{k}\sin\theta\rho(\theta)\frac{\delta B_y}{M-I}.
\label{lmc17}
\end{eqnarray}
By construction,  $\delta\rho_{\|}$ and $\delta\rho_{\perp}$ are orthogonal for the scalar product weighted by $1/\rho$ in the sense that
\begin{eqnarray}
\int \frac{\delta\rho_{\|}\delta\rho_{\perp}}{\rho}\, d\theta=0.
\label{lmc18}
\end{eqnarray}
Indeed, we have
\begin{eqnarray}
\int \frac{\delta\rho_{\|}\delta\rho_{\perp}}{\rho}\, d\theta=\int (\mu+\nu_x\cos\theta+\nu_y\sin\theta)\delta\rho_{\perp}\, d\theta=0,\nonumber\\
\label{lmc19}
\end{eqnarray}
where we have used equations (\ref{lmc7}) and (\ref{lmc7b}) to get the last equality. As a result, we obtain
\begin{eqnarray}
\int \frac{(\delta\rho)^2}{\rho}\, d\theta=\int \frac{(\delta\rho_{\|})^2}{\rho}\, d\theta+\int \frac{(\delta\rho_{\perp})^2}{\rho}\, d\theta.
\label{lmc20}
\end{eqnarray}
Using equations (\ref{lmc17}) and (\ref{lmc15}), we obtain after simplification
\begin{eqnarray}
\int \frac{(\delta\rho_{\|})^2}{\rho}\, d\theta=\frac{2\pi}{k}\lambda'(B)(\delta B_x)^2+\frac{4\pi^2}{k^2}\frac{1}{M-I}(\delta B_y)^2.\nonumber\\
\label{lmc21}
\end{eqnarray}
Therefore, the second order variations of free energy given by equation (\ref{lmc3}) can be written
\begin{eqnarray}
\delta^2F=\frac{\pi}{k}(T\lambda'(B)-1)(\delta B_x)^2\nonumber\\
+\frac{\pi}{k}\left (\frac{2\pi}{k}\frac{T}{M-I}-1\right )(\delta B_y)^2+\frac{1}{2}T\int \frac{(\delta\rho_{\perp})^2}{\rho}\, d\theta.
\label{lmc22} 
\end{eqnarray}
Using identity (\ref{gmc5}), we obtain
\begin{eqnarray}
\delta^2F=\frac{1}{2}F''(B)(\delta B_x)^2\nonumber\\
+\frac{\pi}{k}\left (\frac{2\pi}{k}\frac{T}{M-I}-1\right )(\delta B_y)^2+\frac{1}{2}T\int \frac{(\delta\rho_{\perp})^2}{\rho}\, d\theta.
\label{lmc23}
\end{eqnarray}
For the inhomogeneous phase $B\neq 0$, using equation (\ref{lmc10}) and the self-consistency relation (\ref{se13}), we get
\begin{eqnarray}
I=M-\frac{2\pi}{k}T.
\label{lmc10b}
\end{eqnarray}
In that case, equation (\ref{lmc23}) reduces to
\begin{eqnarray}
\delta^2F=\frac{1}{2}F''(B)(\delta B_x)^2+\frac{1}{2}T\int \frac{(\delta\rho_{\perp})^2}{\rho}\, d\theta.
\label{lmc23b}
\end{eqnarray}
On the other hand, for the homogeneous phase $B=0$, equation
(\ref{lmc10}) leads to
\begin{eqnarray}
I=\frac{M}{2}.
\label{lmc10bb}
\end{eqnarray}
Using equation (\ref{magn30}), 
equation (\ref{lmc23}) reduces to
\begin{eqnarray}
\delta^2F=\frac{1}{2}F''(0)\left\lbrack (\delta B_x)^2+(\delta B_y)^2\right\rbrack+\frac{1}{2}T\int \frac{(\delta\rho_{\perp})^2}{\rho}\, d\theta.\nonumber\\
\label{lmc23bb}
\end{eqnarray}
These relations show that $\rho$ is a local minimum of $F[\rho]$ at fixed mass iff $B$ is a local minimum of $F(B)$. Indeed, if $F''(B)> 0$ then $\delta^2F> 0$ since the last term is positive. On the other hand, if  $F''(B)<0$ it suffices to consider a perturbation of the form (\ref{lmc5}) with  $\delta\rho_{\perp}=0$ and $\delta\rho_{\|}$ given by equation (\ref{lmc17}) to conclude that $\delta^2 F<0$ for this perturbation. This implies that $\rho$ is not a local minimum of $F[\rho]$ since there exists a particular perturbation that decreases the free energy. This is the case for the homogeneous solutions when $T<T_c$ since they are local maxima of $F(B)$. Therefore, (\ref{fr4}) and (\ref{fb12}) are equivalent for local minimization. Combining all our results,  we conclude that the variational problems (\ref{ff1}), (\ref{fr4}) and (\ref{fb12}) are equivalent for local and global minimization:
\begin{eqnarray}
\label{eaqb}
(\ref{ff1}) \Leftrightarrow (\ref{fr4})\Leftrightarrow (\ref{fb12}).
\end{eqnarray}

\section{Microcanonical ensemble}
\label{sec_micro}

\subsection{The functionals $S[f]$ and $S[\rho]$}
\label{sec_sf}

The maximization problem (\ref{sf1}) has several interpretations:

(i) It determines the statistical equilibrium state of the HMF model in the microcanonical ensemble. In that thermodynamical interpretation $S$ is the Boltzmann entropy. The maximization problem (\ref{sf1}) can therefore be interpreted  as a criterion of microcanonical thermodynamical stability.

(ii) It determines a particular steady state of the Vlasov equation that is formally  nonlinearly dynamically stable\footnote{This is a refined condition of formal stability with respect to the usual criterion \cite{holm} since both the mass and the energy are treated here as constraints (see \cite{cd,campac} for a more detailed discussion).}. The maximization problem (\ref{sf1}) can therefore be interpreted as a {\it sufficient} condition of dynamical stability. In that dynamical interpretation, $S$ is a particular Casimir (pseudo entropy). As explained previously, it is convenient to develop a {\it thermodynamical analogy}  \cite{assise} to study this dynamical stability problem and use a common vocabulary.

It is shown in Appendix A.1. of \cite{yukawa} that the solution of (\ref{sf1}) is given by
\begin{eqnarray}
\label{sr2}
f(\theta,v)=\left (\frac{\beta}{2\pi}\right )^{1/2} \, \rho({\theta})\, e^{-\beta\frac{v^2}{2}},
\end{eqnarray}
where $\beta=1/T$ is determined by the energy constraint
\begin{eqnarray}
\label{sr3}
E=\frac{1}{2}MT+W,
\end{eqnarray}
and $\rho(\theta)$ is the solution of
\begin{eqnarray}
\label{sr6}
\max_\rho\left\lbrace S\lbrack \rho\rbrack\, |\, M\lbrack \rho\rbrack=M\right\rbrace,
\end{eqnarray}
where
\begin{eqnarray}
\label{sr4}
S[\rho]=\frac{1}{2}M\ln T-\int \rho\ln\rho\, d\theta.
\end{eqnarray}
Eliminating the temperature thanks to the constraint (\ref{sr3}), we can write the entropy in terms of $\rho$ alone as
\begin{eqnarray}
\label{sr5}
S[\rho]=-\int \rho\ln\rho\, d\theta+\frac{1}{2}M\ln (E-W[\rho]).
\end{eqnarray}
Therefore, the maximization problems (\ref{sf1}) and (\ref{sr6}) are equivalent:
\begin{eqnarray}
\label{eaqc}
(\ref{sf1}) \Leftrightarrow (\ref{sr6}).
\end{eqnarray}
This equivalence holds for global and local maximization \cite{yukawa}: (i) $f(\theta,v)$ is the global maximum of (\ref{sf1}) iff $\rho(\theta)$ is the global maximum of (\ref{sr6}) and (ii) $f(\theta,v)$ is a local maximum of (\ref{sf1}) iff $\rho(\theta)$ is a local maximum of (\ref{sr6}). We are led therefore to considering the maximization problem (\ref{sr6}) which is simpler to study since it involves the density $\rho(\theta)$ instead of the distribution function $f(v,\theta)$.

Before that, let us compare the conditions of stability issued from (\ref{sf1}) and (\ref{sr6}). The critical points of (\ref{sf1}) are given by equations (\ref{se14}), (\ref{se15}), (\ref{se13}) and (\ref{se18}) where $E$ is prescribed. Furthermore, a critical point of $S$ at fixed mass and energy is a (local) maximum iff (see Appendix \ref{sec_deuxm}):
\begin{eqnarray}
\label{sf1b}
\delta^2 S=-\int \frac{(\delta f)^2}{2f}\, d\theta dv-\frac{1}{2}\beta\int\delta\rho\delta\Phi\, d\theta < 0,
\end{eqnarray}
for all perturbations $\delta f$ that do not change the mass and the energy at first order: $\delta M=\delta E=0$. On the other hand, the critical points of (\ref{sr6})  are given by equations (\ref{se15}), (\ref{se13}) and (\ref{se18})  where $E$ is prescribed (see Appendix A.1 of \cite{yukawa}).  Furthermore, a critical point of $S$ at fixed mass is a (local) maximum iff (see Appendix A.1 of \cite{yukawa}):
\begin{eqnarray}
\label{sr8}
\delta^2S=- \int \frac{(\delta\rho)^2}{2\rho}\, d\theta-\frac{1}{2T}\int\delta\rho\delta\Phi\, d\theta\nonumber\\
-\frac{1}{M T^2}\left (\int \Phi\delta\rho\, d\theta\right )^2< 0,
\end{eqnarray}
for all perturbations $\delta\rho$ that conserve mass: $\delta M=0$. This stability criterion is equivalent to the stability criterion (\ref{sf1b}) but it is simpler because it is expressed in terms of the density instead of the distribution function.

{\it Remark 1:} the thermodynamical approach of Kiessling \cite{klast} in the microcanonical ensemble rigorously justifies the  maximization problem (\ref{sf1}).

{\it Remark 2:} comparing the stability criteria (\ref{ff3}) and (\ref{sf1b}), we see that canonical stability implies microcanonical stability in the sense that a (local) minimum of $F$ at fixed mass is necessarily a (local) maximum of $S$ at fixed mass and energy. Indeed, if inequality (\ref{ff3}) is satisfied for all perturbations that conserve mass, then inequality (\ref{sf1b}) is satisfied {\it a fortiori} for all  perturbations that conserve mass {\it and} energy. However, the reciprocal is wrong in case of ensembles inequivalence. Therefore, we just have the implication (\ref{ff1}) $\Rightarrow$ (\ref{sf1}).
This result can also be obtained by comparing the stability criteria (\ref{fr6}) and (\ref{sr8}). Indeed, since the last term in equation (\ref{sr8}) is negative, it is clear that if inequality (\ref{fr6}) is satisfied, then inequality (\ref{sr8}) is automatically satisfied. In general this is not reciprocal and we may have ensembles inequivalence. However, if we consider a spatially homogeneous system for which $\Phi$ is uniform, the last term in equation (\ref{sr8}) vanishes (since the mass is conserved) and the stability criteria (\ref{fr6}) and (\ref{sr8}) coincide. Therefore, for spatially homogeneous systems, we have ensembles equivalence.

{\it Remark 3:} according to the two interpretations of (\ref{sf1}) recalled at the beginning of this section, we note that thermodynamical stability implies dynamical stability (for isothermal distributions). However, the converse is wrong since (\ref{sf1}) provides just a {\it sufficient} condition of dynamical stability. More refined stability criteria are discussed in \cite{cd,campac}.

{\it Remark 4:} the equivalence between (\ref{sf1}) and (\ref{sr6}) can be extended to a larger class of functionals of the form $S[f]=-\frac{1}{q-1}\int (f^q-f)\, d\theta dv$ associated to polytropic distributions. Such functionals (Casimirs) arise in the Vlasov dynamical stability problem. We refer to \cite{cc,cspoly} for a detailed discussion of this equivalence.

\subsection{The function $S(B)$}
\label{sec_sb}

\subsubsection{Global maximization}
\label{sec_gmm}

The equivalence (\ref{eaqc}) is valid for an arbitrary potential of interaction $u({\bf r},{\bf r}')$. Now, for the HMF model, the problem can be simplified further. Indeed, the potential energy is given by equation (\ref{se16}) so that  the energy (\ref{sr3}) and the entropy (\ref{sr5}) can be rewritten
\begin{eqnarray}
\label{sb1}
E=\frac{1}{2}MT-\frac{\pi B^2}{k},
\end{eqnarray}
\begin{eqnarray}
\label{sb2}
S[\rho]=-\int \rho\ln\rho\, d\theta+\frac{1}{2}M\ln \left (E+\frac{\pi B^2}{k}\right ).
\end{eqnarray}
We shall first determine the {\it global} maximum of entropy at fixed mass. To that purpose, we shall reduce the maximization problem (\ref{sr6})  to an equivalent but simpler maximization problem.

To solve the maximization problem  (\ref{sr6}), we proceed in two steps: we first maximize $S[\rho]$ at fixed $M$ {\it and} $B_x$ and $B_y$. Writing the variational problem as
\begin{eqnarray}
\label{sb3}
-\delta \left (\int \rho\ln\rho\, d\theta\right )-\alpha\delta M-\mu_x\delta B_x-\mu_y\delta B_y=0,\nonumber\\
\end{eqnarray}
and proceeding as in Sec. \ref{sec_fb}, we obtain
\begin{eqnarray}
\label{sb4}
\rho_1(\theta)=Ae^{-\lambda\cos(\theta-\phi)},
\end{eqnarray}
where $A$ and $\lambda$ are determined by the constraints $M$ and $B$ through
\begin{equation}
A=\frac{M}{2\pi I_{0}(\lambda)},
\label{sb5}
\end{equation}
and
\begin{equation}
b\equiv \frac{2\pi B}{kM}=\frac{I_{1}(\lambda)}{I_{0}(\lambda)}.
\label{sb6}
\end{equation}
Equation (\ref{sb4}) is the (unique) global maximum of $S[\rho]$ with the previous constraints since $\delta^2 S=-\frac{1}{2}\int \frac{(\delta \rho)^2}{\rho}\, d\theta< 0$ (the constraints are linear in $\rho$ so that their second order variations vanish). Then, we can express the entropy $S$ as a function of $B$ by writing $S(B)\equiv S[\rho_1]$. After straightforward calculations, we obtain
\begin{eqnarray}
S(B)=M\ln I_0(\lambda)-\frac{2\pi B}{k}\lambda+\frac{M}{2}\ln \left (E+\frac{\pi B^2}{k}\right ),\nonumber\\
\label{sb7}
\end{eqnarray}
where $\lambda(B)$ is given by equation (\ref{sb6}). Finally, the maximization problem (\ref{sr6}) is equivalent to the maximization problem
\begin{eqnarray}
\label{sb8}
\max_B\left\lbrace S(B)\right\rbrace,
\end{eqnarray}
in the sense that the solution of (\ref{sr6}) is given by equations (\ref{sb4})-(\ref{sb6})  where $B$ is the solution of (\ref{sb8}). Note that the energy and mass constraints are taken into account implicitly in the variational problem (\ref{sb8}). Therefore, (\ref{sr6}) and (\ref{sb8}) are equivalent for global maximization. However, (\ref{sb8}) is much simpler because, for given $E$ and $M$, we just have to determine the maximum of a {\it function} $S(B)$ instead of the maximum of a functional $S[\rho]$ at fixed mass and energy.

Let us therefore study the function $S(B)$ defined by equations (\ref{sb7}) and (\ref{sb6}). Its first derivative is
\begin{eqnarray}
S'(B)=M\left (\frac{I_1(\lambda)}{I_0(\lambda)}-\frac{2\pi B}{kM}\right )\frac{d\lambda}{dB}-\frac{2\pi }{k}\lambda+\frac{\pi B}{k}\frac{M}{E+\frac{\pi B^2}{k}}.\nonumber\\
\label{gmm1}
\end{eqnarray}
Using the identity $I'_0(x)=I_1(x)$  and the relation (\ref{sb6}), we see that the term in parenthesis vanishes. Then, we get
\begin{eqnarray}
{S'(B)}=\frac{2\pi }{k}\left (\frac{B}{T}-\lambda\right ),
\label{gmm2}
\end{eqnarray}
where $T$ is determined by equation (\ref{sb1}). The critical points of $S(B)$, satisfying $S'(B)=0$, correspond to
\begin{eqnarray}
\lambda=x\equiv \beta B.
\label{gmm3}
\end{eqnarray}
Recalling equations (\ref{sb1}) and (\ref{sb6}), this leads to the self-consistency relations
\begin{equation}
\frac{2\pi B}{kM}=\frac{I_{1}(\beta B)}{I_{0}(\beta B)},
\label{gmm4}
\end{equation}
\begin{eqnarray}
\label{gmm4b}
E=\frac{1}{2}MT-\frac{\pi B^2}{k},
\end{eqnarray}
which determine the magnetization as a function of the energy. This returns the relationships of Sec. \ref{sec_se}.

Now, a critical point of $S(B)$ is a maximum if $S''(B)<0$ and a minimum if $S''(B)>0$. Differentiating  equation (\ref{gmm2}) with respect to $B$, and recalling that the temperature  $T$ is a function of $B$ given by equation (\ref{sb1}), we find that
\begin{eqnarray}
S''(B)=\frac{2\pi }{k}\left (\frac{1}{T}-\frac{d\lambda}{dB}-\frac{4\pi B^2}{kMT^2}\right ).
\label{gmm5}
\end{eqnarray}
Using equation (\ref{gmc5}), we note that
\begin{eqnarray}
S''(B)=-\frac{F''(B)}{T}-\frac{8\pi^2 B^2}{k^2MT^2}.
\label{gmm6}
\end{eqnarray}
Therefore, a critical point of $S(B)$ is a maximum if
\begin{eqnarray}
\frac{d\lambda}{dB}> \frac{1}{T}-\frac{4\pi B^2}{kMT^2},
\label{gmm7}
\end{eqnarray}
and a minimum if the inequality is reversed. Since the last term in equation (\ref{gmm6}) is negative, we recover the fact that canonical stability implies microcanonical stability. Indeed, if the critical point is a minimum of free energy ($F''(B)>0$), then it is a fortiori a maximum of entropy ($S''(B)<0$). Furthermore, we know that the series of equilibria  $E(T)$ is monotonic (see Sec. \ref{sec_se}). Therefore,  the homogeneous solution is a maximum of $S(B)$ for $E>E_c$ since it is a minimum of $F(B)$ for $T>T_c$. On the other hand, the inhomogeneous solution is a maximum of $S(B)$ for $E<E_c$ since it is  a  minimum of $F(B)$ for $T<T_c$. Finally, for the homogeneous solution $B=0$, we have $S''(B)=-F''(B)/T$. Therefore, the homogeneous solution is a local minimum of $S(B)$ for $E<E_c$ since it is a local maximum of $F(B)$ for $T<T_c$. Using the equivalence between (\ref{sr6}) and (\ref{sb8}) for global maximization, we conclude that the maximum entropy state at fixed mass and energy is the homogeneous state for $E>E_c$ and the inhomogeneous state for $E<E_c$. On the other hand, since the ensembles are equivalent for homogeneous solutions (see Remark 2 in Sec. \ref{sec_sf}) and since we have established that the homogeneous solution is a saddle point of free energy for $T<T_c$ (see Sec. \ref{sec_lmc}), we conclude that it is a saddle point of entropy for $E<E_c$. This result is shown by another method  in the next section.

\begin{figure}
\begin{center}
\includegraphics[clip,scale=0.3]{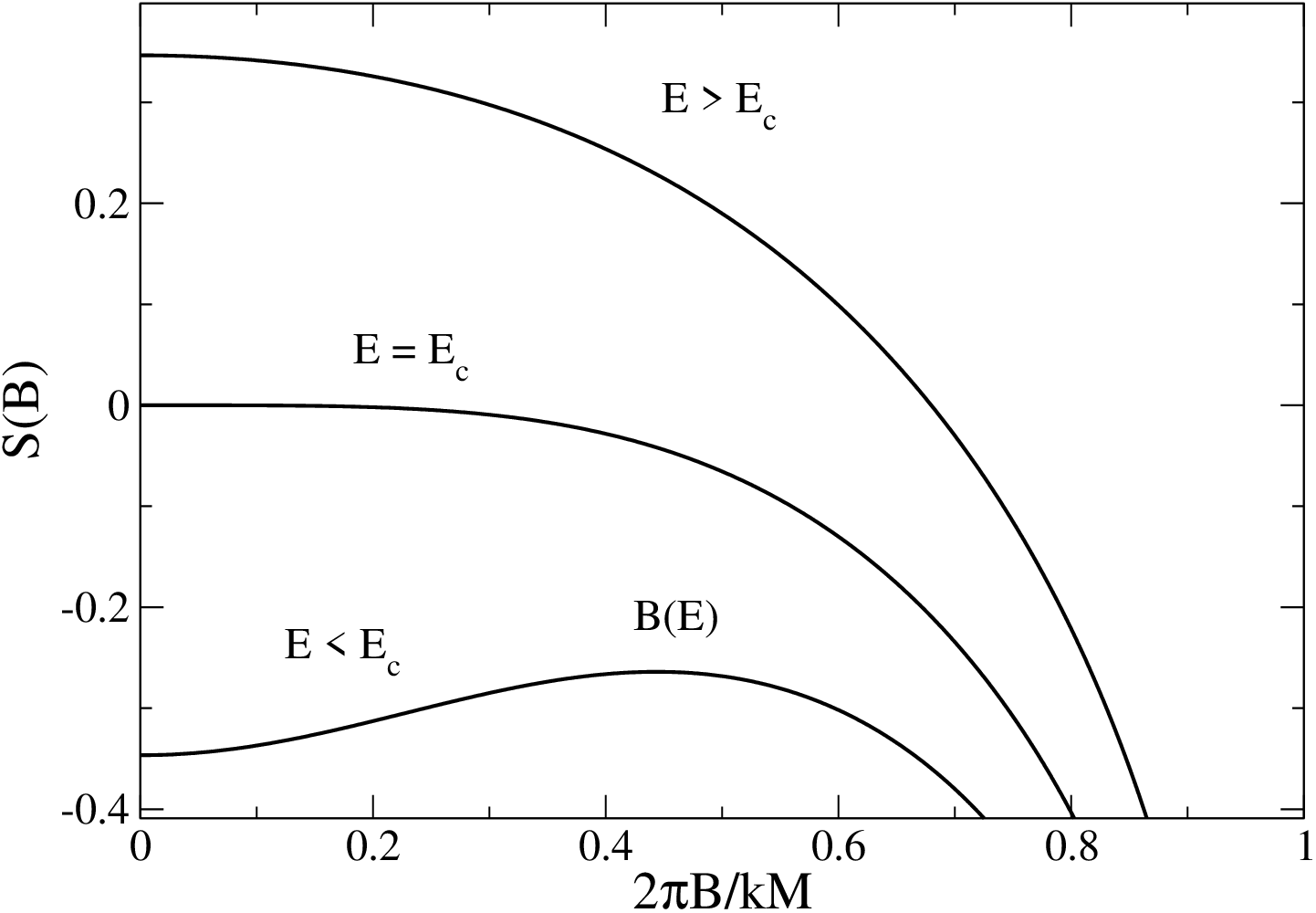}
\caption{Entropy $S(B)$ as a function of magnetization for a given value of energy. For $E>E_c$, this curve has a (unique) global  maximum at $B=0$. For $E<E_c$, this curve has a local minimum at $B=0$ and a global maximum at $B(E)>0$.}
\label{sb}
\end{center}
\end{figure}

To complete our analysis, it can be useful to plot the function $S(B)$ for prescribed mass and energy. Using equations (\ref{sb7}) and (\ref{sb6}), the normalized entropy $s\equiv S/M$ can be expressed in terms of $\lambda$ according to
\begin{eqnarray}
\label{slam}
s(\lambda)=\ln I_0(\lambda)-\lambda\frac{I_1(\lambda)}{I_0(\lambda)}+\frac{1}{2}\ln\left (\epsilon+\frac{2I_{1}(\lambda)^2}{I_{0}(\lambda)^2}\right ).
\end{eqnarray}
Eliminating $\lambda$ between expressions (\ref{slam}) and (\ref{sb6}), we obtain the entropy $s(b)$ as a function of the magnetization $b$ for a fixed value of the energy $E$ (more precisely, for given $\epsilon$, these equations determine $s(b)$  in a parametric form). For $E>E_c$ and $E<E_c$, this function displays the two behaviors described above, as illustrated in Figure \ref{sb}. For $E\rightarrow E_c$ so that $\lambda,b\rightarrow 0$, we find that the entropy takes the approximate form
\begin{eqnarray}
s(b)\simeq \frac{1}{2}\ln\epsilon+\left (\frac{1}{\epsilon}-1\right )b^2-\frac{\epsilon^2+4}{4\epsilon^2}b^4, \quad (\epsilon\rightarrow 1).\nonumber\\
\label{fbzb}
\end{eqnarray}
For $\epsilon<1$, we explicitly check that the maximum satisfying $s'(b)=0$ and $s''(b)<0$ is given by the second relation in equation (\ref{se28}).

{\it Remark:} In Appendix \ref{sec_variance}, we plot the second variations of entropy $S''(B(E))$ (related to the variance of the magnetization) as a function of energy and recover the previous conditions of stability.

\subsubsection{Local maximization}
\label{sec_lmm}

We shall now show that the maximization problems (\ref{sr6}) and (\ref{sb8}) are also equivalent for {\it local} maximization. To that purpose, we shall relate the second order variations of $S[\rho]$ and $S(B)$ by using a suitable decomposition.

A critical point of (\ref{sr6}) is determined by the variational principle
\begin{eqnarray}
\delta S-\alpha \delta M=0,
\label{lmm1}
\end{eqnarray}
where $\alpha$ is a Lagrange multiplier accounting for the conservation of mass. This leads to the distribution (see Appendix A.1. of \cite{yukawa}):
\begin{eqnarray}
\rho=\frac{M}{2\pi I_0(\beta B)} e^{-\beta B\cos\theta},
\label{lmm2}
\end{eqnarray}
where the temperature is determined by the energy according to equation (\ref{sb1}). The magnetization $B$ is obtained by substituting equation (\ref{lmm2}) in equations (\ref{se7})-(\ref{se8}) leading to the self-consistency relation (\ref{se13}). Using equations  (\ref{sr8}) and (\ref{gr3}), a critical point of $S[\rho]$ at fixed mass is a (local) maximum  iff
\begin{eqnarray}
\delta^2 S=-\int \frac{(\delta\rho)^2}{2\rho}\, d\theta+\frac{\pi}{kT} ((\delta B_x)^2+(\delta B_y)^2)\nonumber\\
-\frac{4\pi^2 B^2}{k^2MT^2}(\delta B_x)^2<0,
\label{lmm3}
\end{eqnarray}
for all perturbations $\delta\rho$ that conserve mass: $\int \delta\rho\, d\theta=0$. We note that the second order variations of entropy (\ref{lmm3}) are related to the second order variations of free energy (\ref{lmc3}) by
\begin{eqnarray}
\delta^2 S=-\frac{1}{T}\delta^2F-\frac{4\pi^2 B^2}{k^2MT^2} (\delta B_x)^2.
\label{lmm4}
\end{eqnarray}
Writing the perturbation $\delta\rho$ in the form  (\ref{lmc5}) with equation (\ref{lmc17}) and using expressions (\ref{lmc23b}) and (\ref{lmc23bb}), we obtain for $B\neq 0$:
\begin{eqnarray}
\delta^2 S=-\frac{1}{2}\left (\frac{F''(B)}{T}+\frac{8\pi^2 B^2}{k^2MT^2}\right ) (\delta B_x)^2\nonumber\\
 -\frac{1}{2}\int \frac{(\delta\rho_{\perp})^2}{\rho}\, d\theta.
\label{lmm5b}
\end{eqnarray}
and for $B=0$:
\begin{eqnarray}
\delta^2 S=-\frac{1}{2}\frac{F''(0)}{T} \left\lbrack (\delta B_x)^2+ (\delta B_y)^2\right\rbrack \nonumber\\
 -\frac{1}{2}\int \frac{(\delta\rho_{\perp})^2}{\rho}\, d\theta.
\label{lmm5bb}
\end{eqnarray}
Finally, using identity (\ref{gmm6}), we arrive at
\begin{eqnarray}
\delta^2 S=\frac{1}{2}S''(B)(\delta B_x)^2-\frac{1}{2}\int \frac{(\delta\rho_{\perp})^2}{\rho}\, d\theta,
\label{lmm6}
\end{eqnarray}
for $B\neq 0$ and
\begin{eqnarray}
\delta^2 S=\frac{1}{2}S''(B)\left\lbrack (\delta B_x)^2+ (\delta B_y)^2\right\rbrack -\frac{1}{2}\int \frac{(\delta\rho_{\perp})^2}{\rho}\, d\theta,\nonumber\\
\label{lmm6b}
\end{eqnarray}
for $B=0$.
These relations imply that $\rho$ is a local maximum of $S[\rho]$ at fixed mass iff $B$ is a local maximum of $S(B)$. Indeed, if $S''(B)< 0$, then $\delta^2 S<0$ since the last term is negative. On the other hand, if  $S''(B)>0$, it suffices to consider a perturbation of the form (\ref{lmc5}) with  $\delta\rho_{\perp}=0$ and $\delta\rho_{\|}$ given by equation (\ref{lmc17}) to conclude that $\delta^2 S>0$ for this perturbation. This implies that $\rho$ is not a local maximum of $S[\rho]$ since there exists a particular perturbation that increases the entropy. This is the case for homogeneous solutions when $E<E_c$. Therefore, (\ref{sr6}) and (\ref{sb8}) are equivalent for local maximization. Combining all our results, we conclude that the variational problems (\ref{sf1}), (\ref{sr6}) and (\ref{sb8}) are equivalent for local and global maximization
\begin{eqnarray}
\label{sb9}
(\ref{sf1}) \Leftrightarrow (\ref{sr6}) \Leftrightarrow (\ref{sb8}).
\end{eqnarray}

\section{Conclusion}
\label{sec_conclusion}

In this paper, we have presented a new method to settle the stability of homogeneous and inhomogeneous isothermal distributions in the HMF model. This method starts from general variational principles and transforms them into equivalent but simpler variational principles until a point at which the problem can be easily solved. For isothermal distributions, this method returns, as expected, the same results as those obtained in the past by different procedures \cite{cdr,campac,largedev}, but we would like to emphasize why our approach is interesting and complementary to other methods.

First of all, it is based on general optimization problems: the maximization of $S[f]$ at fixed mass $M$ and energy $E$ or the minimization of $F[f]$ at fixed mass $M$. These optimization problems provide either conditions of {\it thermodynamical stability} (in microcanonical and canonical ensembles respectively) or sufficient conditions of {\it dynamical stability} (more or less refined) with respect to the Vlasov equation. Therefore, our approach allows us to treat thermodynamical and dynamical stability problems with the same formalism. This is not possible if we follow an approach starting directly from the density of states $g(E)$ or the partition function $Z(\beta)$ which only applies to the thermodynamical problem \cite{ms,ar,largedev}. As a consequence, our procedure still works if the Boltzmann functional is replaced by more general functionals of the form $S[f]=-\int C(f)\, d{\bf r}d{\bf v}$, where $C(f)$ is convex. In that case, the maximization of $S$ at fixed $M$ and $E$ or the minimization of $F$ at fixed $M$ provide conditions of formal dynamical stability for arbitrary steady states of the Vlasov equation of the form $f=f(\epsilon)$ with $f'(\epsilon)<0$ \cite{campac}. This dynamical stability problem has been studied recently for polytropic distributions  in \cite{cc}. Our procedure could also be employed for the Lynden-Bell entropy \cite{epjb}, although the calculations would be more complicated. We have chosen to treat here the isothermal case in detail since the calculations are fully analytical. It offers therefore a simple illustration of the method.

Secondly, the optimization problems (\ref{sf1}) and (\ref{ff1}) on which our approach is based are both intuitive and rigorous. The thermodynamical approach is intuitive because the Boltzmann entropy $S[f]$ can be obtained from a simple combinatorial analysis. In that case, the entropy is proportional to the logarithm of the disorder where the disorder measures the number of microstates associated with a given macrostate. Therefore, maximizing entropy $S[f]$ at fixed mass and energy amounts to selecting the most probable macroscopic state consistent with the dynamical constraints\footnote{This result can also be derived from kinetic theory. For isolated systems, the evolution of the distribution function is governed by a kinetic equation that monotonically increases entropy $S[f]$ while conserving mass and energy until the maximum entropy state is reached. Similarly, for systems in contact with a heat bath, the evolution of the distribution function is governed by a kinetic equation that monotonically decreases free energy $F[f]$ while conserving mass until the minimum free energy state is reached \cite{hb}. These H-theorems are another way to justify the  optimization problems (\ref{sf1}) and (\ref{ff1}).}. In the context of the HMF model, this thermodynamical approach has been initiated in \cite{inagaki,cvb} but was not performed to completion since the stability of the inhomogeneous phase was not proven (at least analytically) by this method. This has been done in the present paper. The dynamical approach is also intuitive because we qualitatively understand that the stability of a dynamical system is linked to the fact that the system is in the minimum of a certain ``potential''. For infinite dimensional systems, the nonlinear dynamical stability of a steady state of the Vlasov equation relies on the  energy-Casimir method and its generalizations \cite{campac}. In the context of the HMF model, this approach has been followed in \cite{cvb,yamaguchi,campac}. The optimization problems (\ref{sf1}) and (\ref{ff1}) are also rigorous because they have been given a precise justification by mathematicians. In statistical mechanics, the canonical criterion (\ref{ff1}) has been justified rigorously in \cite{ms} and the microcanonical criterion (\ref{sf1}) in \cite{klast}. Ensembles inequivalence has been formalized in \cite{ellis}.  On the other hand, the formal and nonlinear  dynamical stability of a steady state of the Vlasov equation have been discussed extensively  in the mathematical literature. We refer to \cite{holm} for a survey on the standard energy-Casimir method and to \cite{eht} for refined stability criteria (in the context of 2D turbulence).

Thirdly, our method determines not only the strict caloric curve $\beta(E)$ (corresponding to global maxima of $S$ at fixed $E$ and $M$, or global minima of $F$ at fixed $M$)  but it also provides the whole series of equilibria containing all the critical points of $S$ at fixed $E$ and $M$, or the critical points of $F$ at fixed $M$. In particular, it allows us to determine {\it metastable} states that correspond to local maxima of $S$ at fixed $E$ and $M$, or local minima of $F$ at fixed $M$. There are no such metastable states in the HMF model for isothermal distributions, but they can exist in other situations \cite{staniscia1,cc}. Our approach can be used to determine the sign of the second order variations of the thermodynamical potential in order to settle whether the critical point is a global maximum, a local maximum or a saddle point. In particular, we have been able to relate the second variations of the different functionals in order to show that the equivalence between the optimization problems is not only global but also local.

For these reasons, the approach developed in the present paper is an interesting complement to other methods \cite{cdr,ar,campac,largedev} and it could find application and usefulness in more general situations.

\appendix

\section{Asymptotic expansions}
\label{sec_asy}

In this section, we give the asymptotic expansions of $b(x)$, $\eta(x)$, $\epsilon(x)$, $s(x)$ and $f(x)$ for $x\rightarrow 0$ (corresponding to the bifurcation point) and $x\rightarrow +\infty$ (corresponding to the ground state). 

For $x\rightarrow 0$, we have
\begin{eqnarray}
\label{asy1}
b(x)=\frac{x}{2}-\frac{x^3}{16}+\frac{x^5}{96}+o(x^6),
\end{eqnarray}
\begin{eqnarray}
\label{asy2}
\eta(x)=1+\frac{x^2}{8}-\frac{x^4}{192}+o(x^6),
\end{eqnarray}
\begin{eqnarray}
\label{asy3}
\epsilon(x)=1-\frac{5x^2}{8}+\frac{7x^4}{48}+o(x^6),
\end{eqnarray}
\begin{eqnarray}
\label{asy4}
s(x)=-\frac{5x^2}{16}+\frac{41x^4}{768}+o(x^6),
\end{eqnarray}
\begin{eqnarray}
\label{asy5}
f(x)=1-\frac{5x^4}{128}+o(x^6).
\end{eqnarray}

For $x\rightarrow +\infty$, we have
\begin{eqnarray}
\label{asy6}
b(x)=1-\frac{1}{2x}-\frac{1}{8x^2}+o(x^{-3}),
\end{eqnarray}
\begin{eqnarray}
\label{asy7}
\eta(x)=\frac{x}{2}+\frac{1}{4}+\frac{3}{16x}+\frac{3}{16x^2}+o(x^{-3}),
\end{eqnarray}
\begin{eqnarray}
\label{asy8}
\epsilon(x)=-2+\frac{4}{x}-\frac{1}{x^2}+o(x^{-3}),
\end{eqnarray}
\begin{eqnarray}
\label{asy9}
s(x)=-\ln x+\frac{1}{2}(1-\ln\pi)+\frac{1}{16x^2}+o(x^{-3}),
\end{eqnarray}
\begin{eqnarray}
\label{asy10}
f(x)=-2+\frac{2+2\ln\pi+4\ln x}{x}\nonumber\\
-\frac{\ln\pi+2\ln x}{x^2}+o(x^{-3}).
\end{eqnarray}

\section{Derivation of inequality (\ref{sf1b})}
\label{sec_deuxm}

We consider a perturbation $\delta f$ around a distribution $f$. The exact variations of mass (\ref{gr5}) and energy (\ref{gr6}) for any perturbation are
\begin{eqnarray}
\label{deuxm1}
\Delta M=\int \delta f \, d\theta dv,
\end{eqnarray}
\begin{eqnarray}
\label{deuxm2}
\Delta E=\int \delta f \frac{v^2}{2}\, d\theta dv+\int \Phi\delta\rho\, d\theta+\frac{1}{2}\int\delta\rho\delta\Phi\, d\theta dv.\nonumber\\
\end{eqnarray}
Considering small perturbations $\delta f$, the variations of entropy (\ref{ff2}) up to second order is
\begin{eqnarray}
\label{deuxm3}
\Delta S=-\int (\ln f+1)\delta f \, d\theta dv-\int \frac{(\delta f)^2}{2f}\, d\theta dv.
\end{eqnarray}
Let us now assume that $f$ is a critical points of entropy at fixed mass and energy. It is determined by the variational principle (\ref{se1}), leading to
\begin{eqnarray}
\label{deuxm4}
\ln f+1=-\beta \left (\frac{v^2}{2}+\Phi\right )-\alpha.
\end{eqnarray}
Substituting this relation in equation (\ref{deuxm3}), we obtain
\begin{eqnarray}
\label{deuxm5}
\Delta S=\int \left\lbrack \beta \left (\frac{v^2}{2}+\Phi\right )+\alpha\right \rbrack\delta f \, d\theta dv-\int \frac{(\delta f)^2}{2f}\, d\theta dv.\nonumber\\
\end{eqnarray}
Now, using the conservation of mass and energy $\Delta M=\Delta E=0$, we get
\begin{eqnarray}
\label{deuxm6}
\Delta S=-\int \frac{(\delta f)^2}{2f}\, d\theta dv-\frac{1}{2}\beta\int\delta\rho\delta\Phi\, d\theta dv.
\end{eqnarray}
Requiring that the critical point be a maximum of entropy at fixed mass and energy leads to inequality (\ref{sf1b}). We refer to \cite{campac,ipser} for generalizations of this result to a larger class of functionals in the Vlasov dynamical stability context.

\section{Connection between statistical mechanics and variational principles}
\label{sec_heur}

In this section, we discuss the connection between the statistical
mechanics of systems with long-range interactions and variational
principles.

In the microcanonical ensemble,  the accessible configurations (those having the proper
value of energy) are equiprobable. Therefore, the density probability of the configuration $(\theta_1,v_1,...,\theta_N,v_N)$  is $P_{N}(\theta_{1},v_1,...,\theta_{N},v_N)=\frac{1}{g(E)}\delta(E-H)$ where $g(E)$ is the density of states
\begin{eqnarray}
g(E)=\int \delta(E-H)\, d\theta_1 dv_1...d\theta_N dv_N.
\label{heur1}
\end{eqnarray}
In other words,  $g(E)dE$ gives the number of microstates with energy between $E$ and $E+dE$. 
The entropy is defined by $S(E)=\ln
g(E)$. Let us introduce   the (coarse-grained) one-body
distribution function $f(\theta,v)$. A microstate is determined by the specification of the 
exact positions and
velocities $\lbrace\theta_i,v_i\rbrace$ of the $N$ particles. A
macrostate is determined by  the specification of the density
$\lbrace f(\theta,v)\rbrace$ of particles in each cell $[\theta,\theta+d\theta]\times
[v,v+dv]$ irrespectively of their {\it precise} position in the
cell.  Let us call
$\Omega[f]$ the unconditional number of  microstates 
corresponding to the macrostate $f$.  For $N\rightarrow +\infty$, 
equation (\ref{heur1}) can be formally rewritten
\begin{eqnarray}
g(E)\simeq \int  \Omega[f] \delta(E[f]-E)\delta(M[f]-M)\, {\cal D}f,
\label{heur2}
\end{eqnarray}
where $E[f]$ is the mean field energy (\ref{gr6}) and $M[f]$ is the mass (\ref{gr5}).
The entropy of the macrostate $f$ is defined  by the
Boltzmann formula $S[f]=\ln \Omega[f]$. The
Boltzmann entropy can be obtained by a standard combinatorial analysis
leading, for $N\rightarrow +\infty$,  to equation (\ref{ff2}). Therefore, using  $\Omega[f]= e^{S[f]}$, equation (\ref{heur2}) can be rewritten
\begin{eqnarray}
g(E)\simeq \int  e^{S[f]} \delta(E[f]-E)\delta(M[f]-M)\, {\cal D}f.
\label{heur3}
\end{eqnarray}
The unconditional density probability of the distribution $f$ is $P_0[f]=\frac{1}{\cal A}e^{S[f]}$ (where ${\cal A}$ is the hypervolume of the system in phase space). The microcanonical density probability of the distribution $f$ is $P[f]=\frac{1}{g(E)}e^{S[f]}\delta(E[f]-E)\delta(M[f]-M)$.
 
Integrating over the velocities in equation (\ref{heur1}), a classical calculation leads to 
 \begin{eqnarray}
g(E)=\frac{2\pi^{N/2}}{\Gamma\left (\frac{N}{2}\right )}\int \left\lbrack 2(E-U)\right \rbrack^{\frac{N-2}{2}}\, d\theta_1...d\theta_N,
\label{heur4}
\end{eqnarray}
where $U(\theta_1,...,\theta_N)$ is the potential energy (second term in the r.h.s. of equation (\ref{gr1})).
Let us introduce  the (coarse-grained) one-body density $\rho(\theta)$ and denote by $\Omega[\rho]$ the unconditional number of  microstates $\lbrace \theta_i\rbrace$ corresponding to the macrostate $\rho$. For $N\rightarrow +\infty$,
equation (\ref{heur4}) can be formally rewritten\footnote{Like in the main part of the paper, for the sake of conciseness, we do not explicitly write the constant terms that are independent on $\rho$ in the expression of the entropy (the term in the exponential). They can be restored easily.}
\begin{eqnarray}
g(E)\simeq \int e^{\frac{N}{2}\ln (E-W[\rho])} \Omega[\rho]\delta(M[\rho]-M)\, {\cal D}\rho,
\label{heur5}
\end{eqnarray}
where $W[\rho]$ is the mean field potential energy (second term in the r.h.s. of equation (\ref{gr6})). The unconditional number of microstates $\Omega[\rho]$ can be obtained by  a classical combinatorial analysis leading, for  $N\rightarrow +\infty$, to  $\Omega[\rho]= e^{-\int \rho\ln\rho\, d\theta}$. Therefore, equation (\ref{heur5}) can be rewritten
\begin{eqnarray}
g(E)\simeq \int  e^{S[\rho]} \delta(M[\rho]-M)\, {\cal D}\rho,
\label{heur6}
\end{eqnarray}
where $S[\rho]$ is given by equation (\ref{sr5}). The unconditional density probability of the distribution $\rho$ is $P_0[\rho]=\frac{1}{A}e^{-\int \rho\ln\rho\, d\theta}$ (where $A$ is the volume of the system in physical space). The microcanonical density probability of the distribution $\rho$ is $P[\rho]=\frac{1}{g(E)}e^{S[\rho]}\delta(M[\rho]-M)$.

For the HMF model, let us introduce the magnetization vector
\begin{eqnarray}
B_x=-\frac{k}{2\pi}\sum_{i=1}^{N}\cos\theta_i, \quad B_y=-\frac{k}{2\pi}\sum_{i=1}^{N}\sin\theta_i.
\label{heur7}
\end{eqnarray}
The potential energy can be expressed in terms of the magnetization as
\begin{eqnarray}
U=-\frac{\pi B^2}{k}+\frac{kN}{4\pi}.
\label{heur8}
\end{eqnarray}
Therefore, the density of states (\ref{heur4}) can be rewritten
\begin{eqnarray}
g(E)=\frac{2\pi^{N/2}}{\Gamma\left (\frac{N}{2}\right )}\int \left\lbrack 2\left (E+\frac{\pi B^2}{k}-\frac{kN}{4\pi}\right )\right \rbrack^{\frac{N-2}{2}}\nonumber\\
\times \delta\left (\frac{2\pi B_x}{k}+\sum_{i}\cos\theta_i\right )
\delta\left (\frac{2\pi B_y}{k}+\sum_{i}\sin\theta_i\right )\nonumber\\
\times d\left (\frac{2\pi B_x}{k}\right ) d\left (\frac{2\pi B_y}{k}\right )d\theta_1...d\theta_N,\nonumber\\
\label{heur9}
\end{eqnarray}
or, equivalently,
\begin{eqnarray}
g(E)=\frac{2\pi^{N/2}}{\Gamma\left (\frac{N}{2}\right )}\int  \left\lbrack 2\left (E+\frac{\pi B^2}{k}-\frac{kN}{4\pi}\right )\right \rbrack^{\frac{N-2}{2}} \nonumber\\
\times \Omega\left (\frac{2\pi B_x}{k},\frac{2\pi B_y}{k}\right )    \, d\left (\frac{2\pi B_x}{k}\right ) d\left (\frac{2\pi B_y}{k}\right ),\nonumber\\
\label{heur10}
\end{eqnarray}
where $\Omega({\bf B})$ denotes the unconditional number of microstates $\lbrace \theta_i\rbrace$ corresponding to the macrostate ${\bf B}$.
For $N\rightarrow +\infty$, we have
\begin{eqnarray}
g(E)\simeq \int  e^{\frac{N}{2}\ln \left (E+\frac{\pi B^2}{k}\right )}\Omega({\bf B})\, d{\bf B}.
\label{heur11}
\end{eqnarray}
The computation of $\Omega({\bf B})$ is classical \cite{cdr} and is briefly reproduced in Appendix \ref{sec_magnetization}, with some complements, for the sake of self-consistency. For $N\rightarrow +\infty$, we have
\begin{eqnarray}
\Omega({\bf B})= e^{ -\frac{2\pi B}{k}\lambda+M\ln I_0(\lambda)}.
\label{heur12}
\end{eqnarray}
Therefore, the density of states (\ref{heur11}) can finally be written
\begin{eqnarray}
g(E)\simeq \int  e^{S(B)}\, d{\bf B},
\label{heur13}
\end{eqnarray}
where $S(B)$ is given by equation (\ref{sb7}). The unconditional density probability of the magnetization ${\bf B}$ is $P_0({\bf B})=\frac{1}{A}e^{-\frac{2\pi B}{k}\lambda+M\ln I_0(\lambda)}$. The microcanonical density probability of the magnetization ${\bf B}$ is $P({\bf B})=\frac{1}{g(E)}e^{S(B)}$.

In the canonical ensemble, the density probability of the configuration $(\theta_1,v_1,...,\theta_N,v_N)$  is $P_{N}(\theta_{1},v_1,...,\theta_{N},v_N)=\frac{1}{Z(\beta)}e^{-\beta H}$ where $Z(\beta)$ is the partition function
\begin{eqnarray}
Z(\beta)=\int e^{-\beta H} \, d\theta_1 dv_1...d\theta_N dv_N.
\label{heur14}
\end{eqnarray}
The free energy is defined by $F(\beta)=-\frac{1}{\beta}\ln Z(\beta)$.  Introducing  the unconditional number of microstates $\Omega[f]$ corresponding to the macrostate $f$, we obtain for $N\rightarrow +\infty$:
\begin{eqnarray}
Z(\beta) \simeq  \int e^{-\beta E[f]} \, \Omega[f] \, \delta(M[f]-M)\, {\cal D}f\nonumber\\
\simeq \int  e^{S[f]-\beta E[f]} \, \delta(M[f]-M)\, {\cal D}f\nonumber\\
\simeq \int  e^{-\beta F[f]} \, \delta(M[f]-M)\, {\cal D}f,
\label{heur15}
\end{eqnarray}
where $F[f]=E[f]-TS[f]$ is the Boltzmann free energy (\ref{ff2b}).  The canonical density probability of the distribution $f$ is $P[f]=\frac{1}{Z(\beta)}e^{-\beta F[f]}\delta(M[f]-M)$.

Integrating over the velocities in equation (\ref{heur14}), we get 
\begin{eqnarray}
Z(\beta)=\left (\frac{2\pi}{\beta}\right )^{N/2}\int e^{-\beta U}\, d\theta_1...d\theta_N.
\label{heur15b}
\end{eqnarray}
Introducing  the unconditional number of microstates $\Omega[\rho]$ corresponding to the macrostate $\rho$, we obtain for $N\rightarrow +\infty$:
\begin{eqnarray}
Z(\beta)\simeq \int e^{-\beta W[\rho]}\Omega[\rho]\, \delta(M[\rho]-M)\, {\cal D}\rho\nonumber\\
 \simeq\int e^{-\int\rho\ln\rho\, d\theta-\beta W[\rho]}\, \delta(M[\rho]-M)\, {\cal D}\rho\nonumber\\
\simeq \int e^{-\beta F[\rho]} \, \delta(M[\rho]-M) \, {\cal D}\rho,
\label{heur16}
\end{eqnarray}
where $F[\rho]$ is given by equation (\ref{fr3}).  The canonical density probability of the distribution $\rho$ is $P[\rho]=\frac{1}{Z(\beta)}e^{-\beta F[\rho]}\delta(M[f]-M)$.

For the HMF model for which the potential energy can be expressed in terms of the magnetization, the partition function (\ref{heur15b}) can be rewritten
\begin{eqnarray}
Z(\beta)=\left (\frac{2\pi}{\beta}\right )^{N/2}\int e^{-\beta U} \nonumber\\
\times \delta\left (\frac{2\pi B_x}{k}+\sum_{i}\cos\theta_i\right )
\delta\left (\frac{2\pi B_y}{k}+\sum_{i}\sin\theta_i\right )\nonumber\\
\times d\left (\frac{2\pi B_x}{k}\right ) d\left (\frac{2\pi B_y}{k}\right )d\theta_1...d\theta_N,\nonumber\\
\label{heur16b}
\end{eqnarray}
or, equivalently,
\begin{eqnarray}
Z(\beta)=\left (\frac{2\pi}{\beta}\right )^{\frac{N}{2}}\int e^{\beta\left (\frac{\pi B^2}{k}-\frac{kN}{4\pi}\right )}  \nonumber\\
\times \Omega\left (\frac{2\pi B_x}{k},\frac{2\pi B_y}{k}\right )    \, d\left (\frac{2\pi B_x}{k}\right ) d\left (\frac{2\pi B_y}{k}\right ).
\label{heur16c}
\end{eqnarray}
For $N\rightarrow +\infty$, we get
\begin{eqnarray}
Z(\beta)\simeq \int e^{\beta\frac{\pi B^2}{k}}\Omega({\bf B})\, d{\bf B}\nonumber\\
\simeq \int e^{\beta\frac{\pi B^2}{k}}e^{-\frac{2\pi B}{k}\lambda+M\ln I_0(\lambda)}\, d{\bf B}\nonumber\\
\simeq  \int e^{-\beta F(B)}d{\bf B},
\label{heur16d}
\end{eqnarray}
where $F(B)$ is given by equation (\ref{fb11}).   The canonical density probability of the magnetization ${\bf B}$ is $P({\bf B})=\frac{1}{Z(\beta)}e^{-\beta F(B)}$.

Let us now denote by $\phi$ a generic global variable such as $f(\theta,v)$, $\rho(\theta)$ or $B$. We also recall that for systems with long-range interactions, for which  the mean field approximation is exact in the proper thermodynamic limit $N\rightarrow +\infty$, we have the extensive scalings $S[\phi]=Ns[\phi]$, $E[\phi]=Ne[\phi]$, $F[\phi]=Nf[\phi]$. Accordingly, the preceding results can be formally written
\begin{eqnarray}
g(E)\simeq \int e^{Ns[\phi]}\, \delta(c_{MCE}[\phi]-c_{MCE}) \, {\cal D}\phi,
\label{heur18}
\end{eqnarray}
and
\begin{eqnarray}
Z(\beta)\simeq \int  e^{-\beta Nf[\phi]}\, \delta(c_{CE}[\phi]-c_{CE}) \, {\cal D}\phi,
\label{heur19}
\end{eqnarray}
where the $\delta$-functions take into account the constraints as described above. The microcanonical number of microstates corresponding to the macrostate $\phi$ is $\Omega[\phi]=e^{Ns[\phi]}\, \delta(c_{MCE}[\phi]-c_{MCE})$ and the microcanonical probability of the macrostate $\phi$ is $P[\phi]=\frac{1}{g(E)}e^{Ns[\phi]}\, \delta(c_{MCE}[\phi]-c_{MCE})$. Similarly, the canonical number of microstates corresponding to the macrostate $\phi$ is $\Omega[\phi]=e^{-\beta Nf[\phi]}\, \delta(c_{CE}[\phi]-c_{CE})$ and the canonical probability of the macrostate $\phi$ is $P[\phi]=\frac{1}{Z(\beta)}e^{-\beta Nf[\phi]}\, \delta(c_{CE}[\phi]-c_{CE})$. For $N\rightarrow +\infty$, we can make the saddle point approximation. In the microcanonical ensemble, we obtain
\begin{eqnarray}
g(E)=e^{S(E)}\simeq e^{Ns[\phi_*]},
\label{heur20}
\end{eqnarray}
i.e.
\begin{eqnarray}
\lim_{N\rightarrow +\infty} \frac{1}{N}S(E)=s[\phi_*],
\label{heur21}
\end{eqnarray}
where $\phi_*$ is the solution of the maximization problem
\begin{eqnarray}
\max_{\phi}\lbrace s[\phi]\, |\, E, M\rbrace.
\label{heur22}
\end{eqnarray}
This leads to the variational problems (\ref{sf1}), (\ref{sr6}) and (\ref{sb8}). In the canonical ensemble, we obtain
\begin{eqnarray}
Z(\beta)=e^{-\beta F(\beta)}\simeq e^{-\beta Nf[\phi_*]},
\label{heur23}
\end{eqnarray}
i.e.
\begin{eqnarray}
\lim_{N\rightarrow +\infty} \frac{1}{N}F(\beta)=f[\phi_*],
\label{heur24}
\end{eqnarray}
where $\phi_*$ is the solution of the minimization problem
\begin{eqnarray}
\min_{\phi}\lbrace f[\phi]\, |\, M\rbrace.
\label{heur25}
\end{eqnarray}
This leads to the variational problems (\ref{ff1}), (\ref{fr4}) and (\ref{fb12}).

The preceding discussion shows the connection between the statistical mechanics of systems with long-range interactions (based on the calculation of the density of states and of the partition function) and variational principles (based on the maximization of entropy or minimization of free energy). It also shows how the variational problems (\ref{sf1}), (\ref{sr6}), (\ref{sb8}) and (\ref{ff1}), (\ref{fr4}), (\ref{fb12}) are related to each other. These results can be made rigorous by using the theory of large
deviations. We refer to Ellis \cite{ellisld} for a mathematical presentation
of this theory and to Barr\'e {\it et al.}
\cite{largedev} and Touchette \cite{touchette} for its application to
physical problems. In the present paper, we have considered a different approach. We started from the fundamental variational problems
(\ref{sf1}) and (\ref{ff1}) that can be motivated by a simple combinatorial analysis. This is the historical approach of the problem finding its roots in Boltzmann's work. This is also the traditional approach used by physicists to determine the statistical equilibrium state of self-gravitating
systems (see, e.g., \cite{paddy,ijmpb,antonov,lbw}), two-dimensional point vortices (see,
e.g., \cite{houches,mj}) and the HMF model (see, e.g., \cite{inagaki,cvb}). Therefore, describing the statistical mechanics of systems
with long-range interactions from the fundamental variational problems (\ref{sf1}) and (\ref{ff1}) and
reducing them to simpler but equivalent forms (\ref{sr6}), (\ref{sb8}), (\ref{fr4}) and (\ref{fb12})  as we have done here is an interesting presentation that complements the one followed in \cite{cdr}. A bonus of this approach is that it remains valid when the variational problems (\ref{sf1}) and
(\ref{ff1}) have a dynamical interpretation in relation to the
(formal) nonlinear stability of the system with respect to the Vlasov
equation \cite{campac}. In that case,  $S[f]$ is a Casimir
functional of the form $S[f]=-\int C(f)\, d\theta dv$ (sometimes
called a pseudo entropy) that is more general than the Boltzmann
functional arising in the thermodynamical approach. In particular, in
the dynamical stability problem, the variational problems (\ref{sf1})
and (\ref{ff1})  cannot be obtained from a theory of large deviations since
their physical interpretations are completely different.

{\it Remark 1:} the density of states (\ref{heur1}) and the partition
function (\ref{heur5}) can also be calculated from field theoretical
methods (see e.g.  Horwitz \& Katz \cite{hk} and de Vega \& Sanchez
\cite{vega} for self-gravitating systems and Antoni \& Ruffo \cite{ar} for the
HMF model). These approaches are valuable but they are also
considerably more abstract than the one based on variational
principles.

{\it Remark 2:} the mean field Boltzmann distribution (\ref{se3}) can also
be obtained from the first equation of the Yvon-Bogoliubov-Green (YBG)
hierarchy \cite{hb1}, but the condition of stability (related to
the correlation functions appearing in the next equations of the YBG
hierarchy) is more difficult to obtain than with the approach based on
variational principles.

{\it Remark 3:} for $N\rightarrow +\infty$, the density of states
(resp. partition function) is dominated by the {\it global} maximum of
entropy at fixed mass and energy (resp. minimum of free energy at
fixed mass) according to (\ref{heur20})
(resp. (\ref{heur23})). Nevertheless, {\it local} entropy maxima
(resp. free energy minima), i.e. metastable states, are also fully
relevant because they have very long lifetimes scaling like $e^{N}$
\cite{art,metastable}. This is particularly true in the case of classical
self-gravitating systems for which there is no global entropy maximum
(resp. free energy minimum) \cite{paddy,ijmpb}.

\section{Distribution of the magnetization}
\label{sec_magnetization}

In this Appendix, we determine the distribution of the magnetization
by a direct calculation and show its connection with the entropy $S(B)$ and free energy $F(B)$.  The density probability of the magnetization vector defined by equation (\ref{heur7}) is
\begin{eqnarray}
P\left (\frac{2\pi B_x}{k},\frac{2\pi B_y}{k}\right )=\int\delta\left (\frac{2\pi B_x}{k}+\sum_{i}\cos\theta_i\right )\nonumber\\
\times\delta\left (\frac{2\pi B_y}{k}+\sum_{i}\sin\theta_i\right )P_N(\theta_1,...,\theta_N)\, d\theta_1...d\theta_N.\quad 
\label{magn1}
\end{eqnarray}
In the microcanonical ensemble, the $N$-body distribution function is $P_{N}(\theta_{1},v_1,...,\theta_{N},v_N)=\frac{1}{g(E)}\delta(E-H)$. Integrating over the velocities, a classical calculation gives
\begin{eqnarray}
P_N(\theta_1,...,\theta_N)=\frac{1}{g(E)}\frac{2\pi^{N/2}}{\Gamma\left (\frac{N}{2}\right )}\left\lbrack 2(E-U)\right \rbrack^{\frac{N-2}{2}},
\label{magn2}
\end{eqnarray}
where $U$ is the potential energy (second term in the r.h.s. of equation (\ref{gr1})). In the canonical ensemble, the  $N$-body distribution function is $P_{N}(\theta_{1},v_1,...,\theta_{N},v_N)=\frac{1}{Z(\beta)}e^{-\beta H}$. Integrating over the velocities, we obtain
\begin{eqnarray}
P_N(\theta_1,...,\theta_N)=\frac{1}{Z(\beta)}\left (\frac{2\pi}{\beta}\right )^{N/2}e^{-\beta U}.
\label{magn3}
\end{eqnarray}
Recalling that the potential energy can be expressed in terms of the magnetization according to equation (\ref{heur8}), the density probability of the magnetization in microcanonical and canonical ensembles is given by
\begin{eqnarray}
P\left (\frac{2\pi B_x}{k},\frac{2\pi B_y}{k}\right )=\frac{1}{g(E)}\frac{2\pi^{N/2}}{\Gamma\left (\frac{N}{2}\right )}\nonumber\\
\times \left\lbrack 2\left (E+\frac{\pi B^2}{k}-\frac{kN}{4\pi}\right )\right \rbrack^{\frac{N-2}{2}}\Omega\left (\frac{2\pi B_x}{k},\frac{2\pi B_y}{k}\right ),\nonumber\\
\label{magn4}
\end{eqnarray}
\begin{eqnarray}
P\left (\frac{2\pi B_x}{k},\frac{2\pi B_y}{k}\right )=\frac{1}{Z(\beta)}\left (\frac{2\pi}{\beta}\right )^{N/2} \nonumber\\
\times e^{\beta \left (\frac{\pi B^2}{k}-\frac{kN}{4\pi}\right )} \Omega\left (\frac{2\pi B_x}{k},\frac{2\pi B_y}{k}\right ),
\label{magn5}
\end{eqnarray}
where  
\begin{eqnarray}
\Omega\left (\frac{2\pi B_x}{k},\frac{2\pi B_y}{k}\right )=\int\delta\left (\frac{2\pi B_x}{k}+\sum_{i}\cos\theta_i\right )\nonumber\\
\times\delta\left (\frac{2\pi B_y}{k}+\sum_{i}\sin\theta_i\right )\, d\theta_1...d\theta_N,\nonumber\\
\label{magn6}
\end{eqnarray}
is the unconditional number of microstates with magnetization ${\bf B}$. The calculation of this integral is classical \cite{cdr}. Using the Fourier representation of the $\delta$-function
\begin{eqnarray}
\delta(x)=\int_{-\infty}^{+\infty}e^{iqx}\frac{dq}{2\pi},
\label{magn7}
\end{eqnarray}
we obtain
\begin{eqnarray}
\Omega=(2\pi)^{N-2}\int dq_xdq_y e^{Nh(q_x,q_y)},
\label{magn8}
\end{eqnarray}
where
\begin{eqnarray}
h(q_x,q_y)=i\frac{2\pi}{kM}(q_x B_x+q_y B_y)+\ln J_0(q),
\label{magn9}
\end{eqnarray}
and $q=\sqrt{q_x^2+q_y^2}$. Recalling that $k\sim 1/N$, the function $h$ does not depend on $N$. For $N\rightarrow +\infty$, we can make the saddle point approximation
\begin{eqnarray}
\Omega\sim e^{Nh(q_x^*,q_y^*)}
\label{magn10}
\end{eqnarray}
where $(q_x^*,q_y^*)$ corresponds to the maximum of $h(q_x,q_y)$. The vanishing of $\partial h/\partial q_x$ and   $\partial h/\partial q_y$ leads to
\begin{eqnarray}
i\frac{2\pi}{kM}B_x-\frac{J_1(q)}{J_0(q)}\frac{q_x}{q}=0,
\label{magn11}
\end{eqnarray}
\begin{eqnarray}
i\frac{2\pi}{kM}B_y-\frac{J_1(q)}{J_0(q)}\frac{q_y}{q}=0,
\label{magn12}
\end{eqnarray}
where we have used $J_0'(x)=-J_1(x)$. Setting $q_x=-i\lambda_x$ and $q_y=-i\lambda_y$, we find that $q=i\lambda$ where $\lambda=\sqrt{\lambda_x^2+\lambda_y^2}$. Substituting these expressions in equations (\ref{magn11}) and (\ref{magn12}) and using $J_1(i\lambda)=iI_1(\lambda)$ and $J_0(i\lambda)=I_0(\lambda)$, we find that $\lambda$ is determined by
\begin{eqnarray}
\frac{2\pi B}{kM}=\frac{I_{1}(\lambda)}{I_{0}(\lambda)}.
\label{magn13}
\end{eqnarray}
Then, we get
\begin{eqnarray}
q_x^*=i\lambda\frac{B_x}{B},\quad q_y^*=i\lambda\frac{B_y}{B}.
\label{magn14}
\end{eqnarray}
Substituting these values in equations (\ref{magn9}) and (\ref{magn10}), we finally obtain
\begin{eqnarray}
\Omega({\bf B})\sim e^{N\left\lbrack -\frac{2\pi B}{kM}\lambda+\ln I_0(\lambda)\right \rbrack}.
\label{magn15}
\end{eqnarray}
Therefore, for $N\rightarrow +\infty$, the unconditional density probability of the magnetization, i.e. the one corresponding to a Poissonian (uncorrelated) distribution of angles, is
\begin{eqnarray}
P_0({\bf B})=\frac{1}{A} e^{N\left\lbrack -\frac{2\pi B}{kM}\lambda+\ln I_0(\lambda)\right \rbrack}.
\label{magn16}
\end{eqnarray}
This result can also be obtained from the theory of large deviations by a direct application of the Cramer theorem (see equations (6) and (7) in \cite{largedev}).

We now take into account  the correlations by using the $N$-body distribution functions (\ref{magn2}) and (\ref{magn3}). According to equations (\ref{magn4}) and (\ref{magn15})  the  distribution of the magnetization  in the microcanonical ensemble is given, for $N\rightarrow +\infty$, by
\begin{eqnarray}
P_{MCE}({\bf B})=\frac{1}{g(E)} e^{\frac{M}{2}\ln \left (E+\frac{\pi B^2}{k}\right )-\frac{2\pi B}{k}\lambda+M\ln I_0(\lambda)}.\nonumber\\
\label{magn17}
\end{eqnarray}
It can be written
\begin{eqnarray}
P_{MCE}({\bf B})=\frac{1}{g(E)}e^{S(B)},
\label{magn18}
\end{eqnarray}
where $S(B)$ is the entropy defined in equation (\ref{sb7}). According to equations (\ref{magn5}) and (\ref{magn15}), the  distribution of the magnetization  in the canonical ensemble is given, for $N\rightarrow +\infty$, by
\begin{eqnarray}
P_{CE}({\bf B})=\frac{1}{Z(\beta)} e^{\frac{\beta \pi B^2}{k}-\frac{2\pi B}{k}\lambda+M\ln I_0(\lambda)}.
\label{magn19}
\end{eqnarray}
It can be written
\begin{eqnarray}
P_{CE}({\bf B})=\frac{1}{Z(\beta)}e^{-\beta F(B)},
\label{magn20}
\end{eqnarray}
where $F(B)$ is the free energy defined in equation (\ref{fb11}).

\section{Variance of the magnetization}
\label{sec_variance}

In this Appendix, we show the connection between the variance of the magnetization in canonical and microcanonical ensembles and the second order derivatives of entropy and free energy.

The unconditional density probability of the magnetization (\ref{magn16}) can be written
\begin{eqnarray}
P_0({\bf B})=\frac{1}{A} e^{Nh(B)},
\label{magn21}
\end{eqnarray}
where we have introduced the function
\begin{eqnarray}
h(B)=-\frac{2\pi B}{kM}\lambda+\ln I_0(\lambda),
\label{magn22}
\end{eqnarray}
where $\lambda(B)$ is defined by equation (\ref{magn13}).
Note that $h(B)$ can be interpreted as the entropy of the magnetization for a Poissonian distribution of angles. For $N\rightarrow +\infty$, the distribution is strongly peaked around its maximum. The most probable value of the magnetization $B$ corresponds to the maximum of $h(B)$. Using equation (\ref{magn13}), the first derivative of $h(B)$ is
\begin{eqnarray}
h'(B)=-\frac{2\pi \lambda}{kM},
\label{magn23}
\end{eqnarray}
so that the most probable value of the magnetization is $\lambda=B=0$ (we will see that it corresponds indeed to a maximum). Let us expand $h(B)$ around its maximum $h(0)=0$. The second derivative of $h(B)$ is
\begin{eqnarray}
h''(B)=-\frac{2\pi}{kM}\frac{d\lambda}{dB}.
\label{magn24}
\end{eqnarray}
Using identity (\ref{lmc13b}), we obtain
\begin{eqnarray}
h''(0)=-\frac{8\pi^2}{k^2M^2}<0,
\label{magn25}
\end{eqnarray}
justifying that $B=0$ really is the maximum of $h(B)$. Therefore, for $\sqrt{N}B\sim 1$, the distribution of the magnetization is the Gaussian:
\begin{eqnarray}
P_0({\bf B})=\frac{4\pi}{k^2M} e^{-\frac{4\pi^2B^2}{k^2 M}}.
\label{magn26}
\end{eqnarray}
This result can be directly obtained from the central limit theorem. The variance of the unconditional  distribution of magnetization is
\begin{eqnarray}
\langle B^2\rangle_{0}=\frac{k^2 M}{4\pi^2}.
\label{magn27}
\end{eqnarray}

Let us now consider the distribution of the magnetization in the canonical ensemble given by equation (\ref{magn20}). The most probable value of $B$ corresponds to the minimum of free energy $F(B)$ as studied in section \ref{sec_fb}. The vanishing of $F'(B)$ leads to equation (\ref{gmc4}). The second derivative of $F(B)$ at the extremum point is given by equation (\ref{gmc5}) where $\lambda'(B)$ is given by equation (\ref{lmc13b}).

In the homogeneous phase $B=0$, the variance of the magnetization is
\begin{eqnarray}
\langle B^2\rangle_{CE}=\frac{2}{\beta F''(0)}.
\label{magn28}
\end{eqnarray}
Using equation (\ref{lmc13b}), we obtain
\begin{eqnarray}
\lambda'(0)=\frac{4\pi}{kM}.
\label{magn29}
\end{eqnarray}
According to equation (\ref{gmc5}), we have
\begin{eqnarray}
\beta F''(0)=\frac{8\pi^2}{k^2M}\left (1-\frac{T_c}{T}\right ).
\label{magn30}
\end{eqnarray}
Therefore, using equation (\ref{magn28}), we obtain
\begin{eqnarray}
\langle B^2\rangle_{CE}=\frac{k^2M}{4\pi^2}\frac{1}{1-T_c/T}.
\label{magn31}
\end{eqnarray}
We first note that, according to equation (\ref{magn30}),  the homogeneous phase is a minimum of free energy ($F''(0)>0$) for  $T>T_c$ and a maximum of free energy ($F''(0)<0$) for  $T<T_c$ in agreement with the graphical construction of section \ref{sec_fb}. On the other hand, for $T\rightarrow +\infty$, i.e. $\beta=0$, the distribution of angles becomes uniformly distributed and uncorrelated (see equation (\ref{magn3})) so that  we recover the result (\ref{magn27}) valid for a Poissonian distribution. Finally, we note that the variance diverges for $T\rightarrow T_c^+$. This result was previously obtained in \cite{cvb,hb1} (expressed in terms of the variance of the force $\langle F^2\rangle=\langle B^2\rangle/2$) from the second equation of the YBG hierarchy.

In the inhomogeneous  phase $B\neq 0$, the variance of the magnetization is
\begin{eqnarray}
\langle (\Delta B)^2\rangle_{CE}=\frac{1}{\beta F''(B)}.
\label{magn32}
\end{eqnarray}
Using equations (\ref{lmc13b}) and (\ref{gmc4}), we obtain
\begin{eqnarray}
\frac{1}{\lambda'(B)}=\frac{kM}{2\pi}-T-\frac{2\pi B^2}{kM}.
\label{magn33}
\end{eqnarray}
According to equation (\ref{gmc5}), we have
\begin{eqnarray}
\beta F''(B)=\frac{8\pi^2}{k^2M}\left (\frac{1}{2-\frac{T}{T_c}-\frac{8\pi^2 B^2}{k^2 M^2}}-\frac{T_c}{T}\right ).
\label{magn34}
\end{eqnarray}
Therefore, the variance of the magnetization is given by equations (\ref{magn32}) and  (\ref{magn34}) where $B(T)$ is given by equation ({\ref{gmc4}). Figure \ref{varianceeta} shows that $F''(B(T))$ is always positive so that the inhomogeneous phase is always a minimum of free energy in agreement  with the graphical construction of section \ref{sec_fb}.

Let us finally consider the distribution of the magnetization in the microcanonical ensemble given by equation (\ref{magn18}). The most probable value of $B$ corresponds to the maximum of entropy $S(B)$ as studied in section \ref{sec_gmm}. The vanishing of $S'(B)$ leads to equation (\ref{gmm4}) with equation (\ref{gmm4b}). The second derivative of $S(B)$ is related to the second derivative of free energy by equation (\ref{gmm6}).

In the homogeneous phase $B=0$, the variance of the magnetization is
\begin{eqnarray}
\langle B^2\rangle_{MCE}=-\frac{2}{S''(0)}.
\label{magn35}
\end{eqnarray}
Using equations (\ref{gmm6}) and (\ref{magn30}), we obtain
\begin{eqnarray}
S''(0)=-\beta F''(0)=-\frac{8\pi^2}{k^2M}\left (1-\frac{T_c}{T}\right ).
\label{magn36}
\end{eqnarray}
Using equation (\ref{gmm4b}) with $B=0$, this can be expressed in terms of the energy as
\begin{eqnarray}
S''(0)=-\frac{8\pi^2}{k^2M}\left (1-\frac{E_c}{E}\right ).
\label{magn37}
\end{eqnarray}
Therefore, using equation (\ref{magn35}), we obtain
\begin{eqnarray}
\langle B^2\rangle_{MCE}=\frac{k^2M}{4\pi^2}\frac{1}{1-E_c/E}.
\label{magn38}
\end{eqnarray}
We note that, according to equation (\ref{magn37}),  the homogeneous phase is a maximum of entropy ($S''(0)<0$) for  $E>E_c$ and a minimum  of entropy ($S''(0)>0$) for  $E<E_c$ in agreement with the discussion of section \ref{sec_gmm}. On the other hand, the variances of the magnetization in canonical and microcanonical ensembles coincide: $\langle B^2\rangle_{CE}=\langle B^2\rangle_{MCE}$.

In the inhomogeneous  phase $B\neq 0$, the variance of the magnetization is
\begin{eqnarray}
\langle (\Delta B)^2\rangle_{MCE}=-\frac{1}{S''(E)}.
\label{magn39}
\end{eqnarray}
Using equations (\ref{gmm6}) and (\ref{magn34}), we obtain
\begin{eqnarray}
S''(B)=-\frac{8\pi^2}{k^2M}\left (\frac{1}{2-\frac{T}{T_c}-\frac{8\pi^2 B^2}{k^2 M^2}}-\frac{T_c}{T}+\frac{B^2}{T^2}\right ).\nonumber\\
\label{magn40}
\end{eqnarray}
Therefore, the variance of the magnetization is given by equations (\ref{magn39}),  (\ref{magn40}) where $B(T)$ is given by equation ({\ref{gmm4}). It can be expressed in terms of the energy by using equation (\ref{gmm4b}). Figure \ref{varianceepsilon} shows that $S''(B)$ is always negative so that the inhomogeneous phase is always a maximum of entropy in agreement  with the results of section \ref{sec_gmm}. The variances of the magnetization in canonical and microcanonical ensembles do {\it not} coincide in the inhomogeneous phase: $\langle (\Delta B)^2\rangle_{CE}\neq \langle (\Delta B)^2\rangle_{MCE}$.

\begin{figure}
\begin{center}
\includegraphics[clip,scale=0.3]{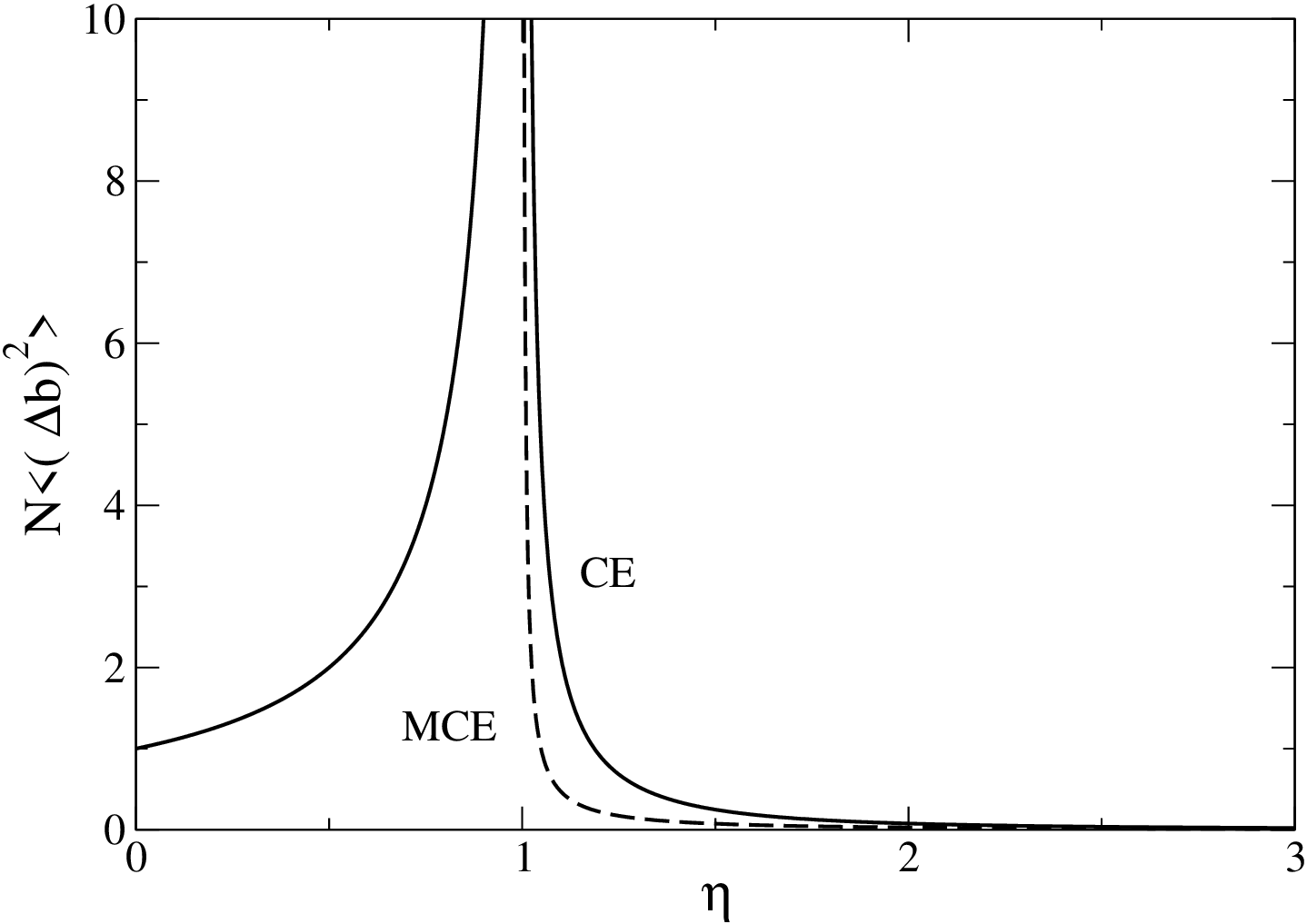}
\caption{Variance of the magnetization as a function of the inverse temperature. We have represented the variance in the canonical (full line) and microcanonical (dashed line) ensembles.}
\label{varianceeta}
\end{center}
\end{figure}

\begin{figure}
\begin{center}
\includegraphics[clip,scale=0.3]{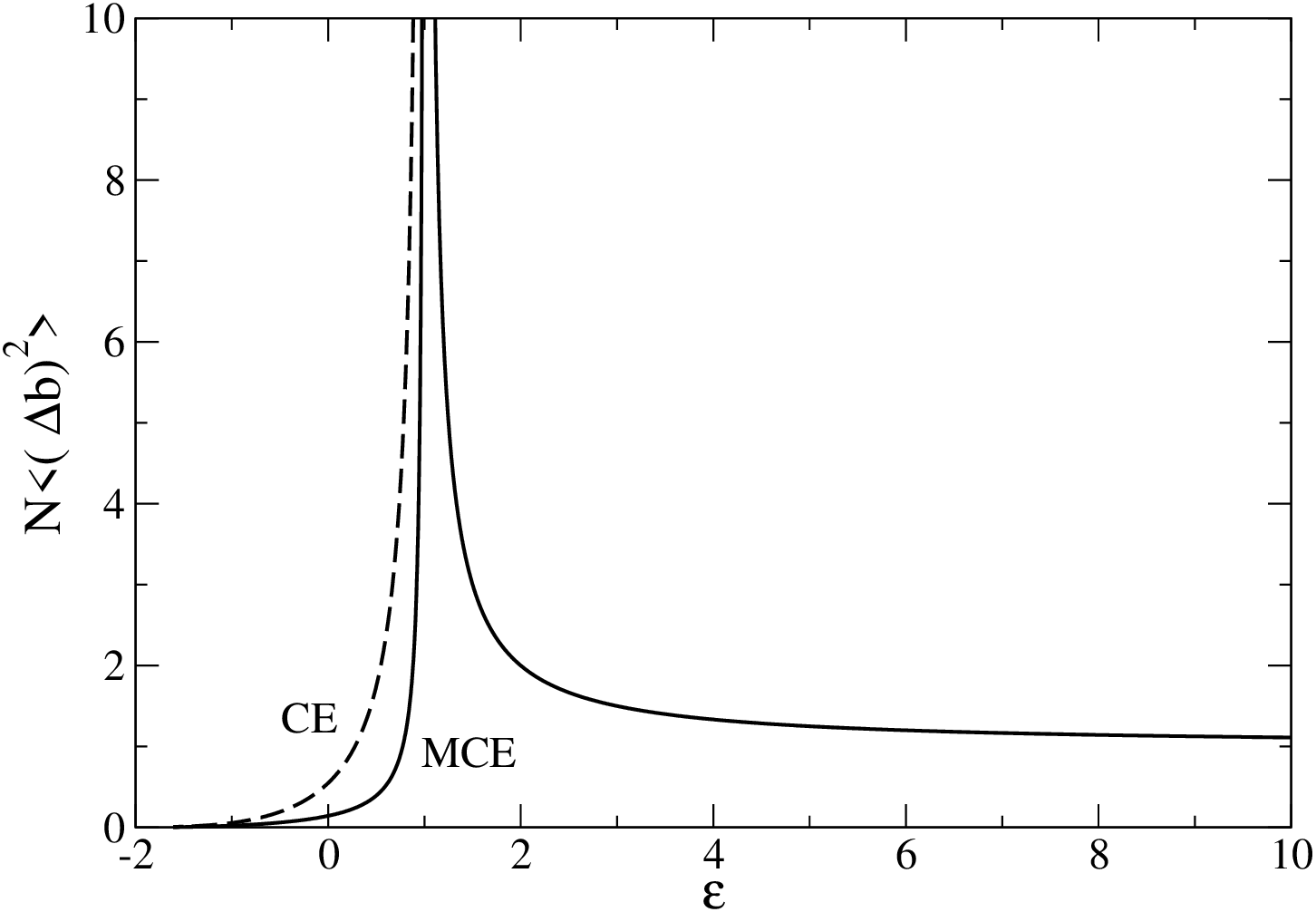}
\caption{Variance of the magnetization as a function of the energy. We have represented the variance in the microcanonical (full line) and canonical (dashed line) ensembles.}
\label{varianceepsilon}
\end{center}
\end{figure}

In order to represent these results graphically (see Figures
\ref{varianceeta} and \ref{varianceepsilon}), it is convenient to
introduce the dimensionless variables defined in section
\ref{sec_se}. The variance of the magnetization in the homogeneous
phase is
\begin{eqnarray}
N\langle b^2\rangle=\frac{1}{1-\eta}=\frac{1}{1-1/\epsilon},
\label{magn41}
\end{eqnarray}
both in canonical and microcanonical ensembles. The variance of the magnetization in the inhomogeneous phase is 
\begin{eqnarray}
N\langle (\Delta b)^2\rangle_{CE}=\frac{1}{\frac{1}{1-\frac{1}{2\eta}-b^2}-2\eta},
\label{magn42}
\end{eqnarray}
in the canonical ensemble and 
\begin{eqnarray}
N\langle (\Delta b)^2\rangle_{MCE}=\frac{1}{\frac{1}{1-\frac{1}{2\eta}-b^2}-2\eta+8b^2\eta^2},
\label{magn43}
\end{eqnarray}
in the microcanonical ensemble. In the microcanonical ensemble, it can be expressed in terms of the energy, using $\epsilon=1/\eta-2b^2$. Using the asymptotic expansions of Appendix \ref{sec_asy}, we obtain close to the bifurcation point $(\epsilon,\eta)\rightarrow (1^{-},1^{+})$:
\begin{eqnarray}
N\langle (\Delta b)^2\rangle_{CE}\sim\frac{1}{4(\eta-1)}\sim\frac{5}{4(1-\epsilon)},
\label{magn44}
\end{eqnarray}
\begin{eqnarray}
N\langle (\Delta b)^2\rangle_{MCE}\sim\frac{1}{20(\eta-1)}\sim\frac{1}{4(1-\epsilon)}.
\label{magn45}
\end{eqnarray}
On the other hand, close to the ground state  $(\epsilon,\eta)\rightarrow (-2^{+},+\infty)$:
\begin{eqnarray}
N\langle (\Delta b)^2\rangle_{CE}\sim\frac{1}{8\eta^2}\sim\frac{(\epsilon+2)^2}{32},
\label{magn46}
\end{eqnarray}
\begin{eqnarray}
N\langle (\Delta b)^2\rangle_{CE}\sim\frac{1}{16\eta^2}\sim\frac{(\epsilon+2)^2}{64}.
\label{magn47}
\end{eqnarray}

We emphasize that althought the ensembles are equivalent regarding the
caloric curve $\beta(E)$, the variance of the magnetization in the
inhomogeneous phase differs in the two ensembles. Therefore, numerical
simulations of the isolated HMF model (microcanonical ensemble) and of
the dissipative BMF model (canonical ensemble) should lead to
different values of $\langle (\Delta B)^2\rangle$.


\begin{thebibliography}{}


\bibitem{houches} {\small {\it Dynamics and thermodynamics of systems
with long range interactions}, edited by T. Dauxois {\it et al.},
Lecture Notes in Physics {\bf 602}, (Springer, 2002)}
\bibitem{assisebook} {\small {\it Dynamics and thermodynamics of systems
with long range interactions: Theory and experiments}, edited by
A. Campa {\it et al.}, AIP Conf. Proc. {\bf 970} (AIP, 2008). }
\bibitem{oxford}  {\small  {\it Long-Range Interacting Systems}, edited by T. Dauxois, S. Ruffo and L. Cugliandolo, Les Houches Summer School 2008, (Oxford: Oxford University Press, 2009)}
\bibitem{cdr}  {\small A. Campa, T. Dauxois, S. Ruffo,   Physics Reports {\bf 480}, 57 (2009)}
\bibitem{lb}  {\small  D. Lynden-Bell, Mon. Not. R. astr. Soc.  {\bf 136}, 101 (1967)}
\bibitem{ellis} {\small  R. Ellis, K. Haven, B. Turkington, J. Stat. Phys.  {\bf 101}, 999 (2000)}
\bibitem{bb}  {\small F. Bouchet, J. Barr\'e, J. Stat. Phys. {\bf 118}, 1073 (2005)}
\bibitem{paddy}  {\small T. Padmanabhan, Phys. Rep. {\bf 188}, 285 (1990)}
\bibitem{katzrev}  {\small J. Katz, Found. Phys. {\bf 33}, 223 (2003)}
\bibitem{ijmpb}  {\small P.H. Chavanis, Int J. Mod. Phys. B {\bf 20}, 3113 (2006)}
\bibitem{ms}  {\small J. Messer, H. Spohn, J. Stat. Phys. {\bf 29}, 561 (1982)}
\bibitem{kk}  {\small T. Konishi, K. Kaneko, J. Phys. A {\bf 25}, 6283 (1992)}
\bibitem{ik}  {\small S. Inagaki, T. Konishi, Publ. Astron. Soc. Japan {\bf 45}, 733 (1993)}
\bibitem{inagaki}  {\small S. Inagaki, Prog. Theor. Phys. {\bf 90}, 557 (1993)}
\bibitem{pichon}  {\small  C. Pichon, PhD thesis, Cambridge (1994)}
\bibitem{ar}  {\small  M. Antoni, S.  Ruffo, Phys. Rev. E {\bf  52}, 2361 (1995)  }
\bibitem{cvb}  {\small P.H. Chavanis, J. Vatteville, F. Bouchet, Eur. Phys. J. B {\bf 46}, 61 (2005)}
\bibitem{epjb}  {\small P.H. Chavanis, Eur. Phys. J. B {\bf 53}, 487 (2006)}
\bibitem{precommun}  {\small  A. Antoniazzi, D. Fanelli, J. Barr\'e, P.H. Chavanis, T. Dauxois, S. Ruffo, Phys. Rev. E {\bf 75}, 011112 (2007) }
\bibitem{anto}  {\small A. Antoniazzi, D. Fanelli, S. Ruffo, Y. Yamaguchi, Phys. Rev. Lett. {\bf 99}, 040601 (2007)}
\bibitem{proc}  {\small P.H. Chavanis, G. De Ninno, D. Fanelli, S. Ruffo, {\it Out of equilibrium phase transitions in mean field Hamiltonian dynamics} in {Chaos, Complexity and Transport: Theory and Applications}, edited by C. Chandre, X. Leoncini, G. Zaslavsky  (World Scientific 2008)}
\bibitem{staniscia1}  {\small F. Staniscia, P.H. Chavanis, G. de Ninno, D. Fanelli, Phys. Rev. E {\bf 80}, 021138 (2009)}
\bibitem{incomplete}  {\small  P.H. Chavanis, Physica A {\bf 365}, 102 (2006)}
\bibitem{cc}  {\small P.H. Chavanis, A. Campa, to appear in EPJB [arXiv:1001.2109]}
\bibitem{yamaguchi}  {\small Y.Y. Yamaguchi, J. Barr\'e, F. Bouchet, T. Dauxois, S. Ruffo,  Physica A  {\bf 337}, 36 (2004)}
\bibitem{choi}  {\small M.Y. Choi, J. Choi, Phys. Rev. Lett. {\bf 91}, 124101 (2003)}
\bibitem{jain}  {\small K. Jain, F. Bouchet, D. Mukamel, J. Stat. Mech. P11008 (2007)}
\bibitem{cd}  {\small P.H. Chavanis, L. Delfini, Eur. Phys. J. B {\bf 69}, 389 (2009)}
\bibitem{campac}  {\small A. Campa, P.H. Chavanis, J. Stat. Mech. P06001 (2010)}	
\bibitem{largedev}  {\small J. Barr\'e, F. Bouchet, T. Dauxois, S. Ruffo, J. Stat. Phys. {\bf 119}, 677 (2005)}
\bibitem{katz}  {\small J. Katz, Mon. Not. R. astr. Soc. {\bf  183}, 765 (1978) }
\bibitem{lrt}  {\small V. Latora, A. Rapisarda, C. Tsallis,  Physica A {\bf 305}, 129 (2002)}
\bibitem{holm}  {\small D.D. Holm, J.E. Marsden, T. Ratiu, A. Weinstein, Phys. Rep. {\bf 123}, 1 (1985)}
\bibitem{assise}  {\small P.H. Chavanis, AIP Conf. Proc. {\bf 970}, 39 (2008)}
\bibitem{bt}  {\small  J. Binney, S. Tremaine,
{\it Galactic Dynamics} (Princeton Series in Astrophysics, 1987)}
\bibitem{aaantonov}  {\small  P.H. Chavanis, A\&A {\bf 451}, 109 (2006)}
\bibitem{yukawa}  {\small P.H. Chavanis, L. Delfini, Phys. Rev. E {\bf 81}, 051103 (2010)}	
\bibitem{nfp}  {\small P.H. Chavanis, Eur. Phys. J. B {\bf 62}, 179 (2008)}
\bibitem{nyquistgrav}  {\small P.H. Chavanis, [arXiv:1002.0291]}
\bibitem{bouchet} {\small F. Bouchet, Physica D {\bf 237}, 1978 (2008)}
\bibitem{euler} {\small P.H. Chavanis, Eur. Phys. J. B {\bf 70}, 73 (2009)}
\bibitem{klast} {\small M. Kiessling, Rev. Math. Phys. {\bf 21}, 1145 (2009)}	\bibitem{cspoly}  {\small P.H. Chavanis, C. Sire, Phys. Rev. E {\bf 69}, 016116 (2004)}
\bibitem{hb}  {\small P.H. Chavanis, Physica A {\bf 361}, 81 (2006)}
\bibitem{eht} {\small R.S. Ellis, K. Haven, B. Turkington, Nonlinearity {\bf 15}, 239 (2002).}
\bibitem{ipser}  {\small J.R. Ipser, G. Horwitz, Astrophys. J. {\bf 232}, 863 (1979)}
\bibitem{ellisld}  {\small R.S. Ellis, {\it Entropy, Large Deviations, and Statistical Mechanics } (Springer-Verlag, New York 1985)}
\bibitem{touchette}  {\small H. Touchette, Phys. Rep. {\bf 478}, 1 (2009)}
\bibitem{antonov} {\small V.A. Antonov, Vest. Leningr. Gos. Univ. {\bf 7}, 135 (1962).}
\bibitem{lbw} {\small D. Lynden-Bell, R. Wood, Mon. not. R. astron. Soc. {\bf 138}, 495 (1968)}	
\bibitem{mj}  {\small  D. Montgomery, G. Joyce,  Phys. Fluids {\bf 17}, 1139  (1974)}
\bibitem{hk}  {\small G. Horwitz, J. Katz, Astrophys. J. {\bf 211}, 226 (1977)}
\bibitem{vega} {\small de Vega, H.J., Sanchez, N. : Nucl. Phys. B {\bf 625}, 409 (2002)}
\bibitem{hb1}  {\small P.H. Chavanis, Physica A  {\bf 361}, 55 (2006) }
\bibitem{art}  {\small  M. Antoni, S. Ruffo, A. Torcini, Europhys. Lett., {\bf  66}, 645 (2004)}
\bibitem{metastable}  {\small  P.H. Chavanis, Astron. Astrophys. {\bf  432}, 117 (2005)}



\end{thebibliography}
\end{document}